\documentclass[a4paper, 11pt]{article}
\usepackage[pdftex, a4paper]{geometry}
\usepackage{graphicx}				
				
\usepackage{fancyhdr}
\usepackage{amsmath}
\usepackage{amsfonts}
\usepackage{dsfont}
\usepackage{mathrsfs}
\usepackage{amssymb}
\usepackage{bbm}
\usepackage{epsfig}
\usepackage[english]{babel}
\usepackage[all]{xy}
\usepackage{subfigure}
\usepackage{color}

\usepackage{natbib}
 \bibpunct[, ]{(}{)}{,}{a}{}{,}%

\newtheorem{theorem}{Theorem}
\newtheorem{definition}{Definition}

\newtheorem{lemma}{Lemma}
\newtheorem{remark}{Remark}
\newtheorem{proposition}{Proposition}
\newtheorem{assumption}{Assumption}
\newtheorem{example}{Example}

\begin{document}
\title{\bf The Evolution of Beliefs over Signed Social Networks}
\date{}

\author{Guodong Shi,  Alexandre Proutiere, \\ Mikael Johansson,  John S. Baras,   and Karl Henrik Johansson}
\maketitle

\begin{abstract}
We study the evolution of opinions (or beliefs) over a  social network modeled as a signed graph. The sign attached to an edge in this graph characterizes whether the corresponding individuals or end nodes are friends (positive links) or  enemies  (negative links). Pairs of nodes are randomly selected to interact over time, and when two nodes interact, each of them updates its opinion based on the opinion of the other node and the sign of the corresponding link. This model generalizes DeGroot model to account for negative links: when two enemies interact, their opinions go in opposite directions. We provide conditions for convergence and divergence in expectation, in mean-square, and in almost sure sense, and exhibit  phase transition phenomena for these notions of convergence depending on the parameters of the opinion update model and on the structure of the underlying graph. We establish a {\it no-survivor} theorem, stating that the difference in opinions of any two nodes diverges whenever opinions in the network diverge as a whole. We also prove a {\it live-or-die} lemma, indicating that almost surely, the opinions either converge to an agreement or diverge. Finally, we extend our analysis to cases where opinions have hard lower and upper limits. In these cases, we study when and how opinions may become asymptotically clustered to the belief boundaries, and highlight the crucial influence  of  (strong or weak) structural balance of the underlying network on this clustering phenomenon.
\end{abstract}

\section{Introduction}

\subsection{Motivation}

We all form opinions about economical, political, and social events that take place in society. These opinions can be binary (e.g., whether one supports a candidate in an election or not) or continuous (to what degree one expects a prosperous future economy). Our opinions are revised when we interact with each other over various  social networks.
Characterizing the evolution of opinions, and understanding the dynamic and asymptotic behavior of the social belief,  are fundamental challenges in the theoretical study of social networks.

Building a good model on how individuals interact and influence each other is essential for studying opinion dynamics. In interaction models, it is natural that a trusted friend  should have a different influence on the opinion formation  than a dubious stranger. The observation that sentiment influences opinions can be traced back to the 1940's when \cite{heider46} introduced the theory of signed social networks, where each interaction link in the social network is associated with a sign (positive or negative) indicating whether  two  individuals  are friends or enemies. Efforts to understand the 
properties of signed social networks have led to the development of structural balance theory, with seminal contributions by \cite{harary56} and \cite{davis63,davis67}. A fundamental insight from these studies, formalized in Harary's theorem~\cite{harary53}, is that local structural properties imply hard global constraints on the social network formation.

In this paper, we attempt to model the evolution of opinions in signed social networks when local hostile or antagonistic relations influence the global social belief. The relative strengths and structures of positive and negative relations are shown to have an essential effect on opinion convergence. In some cases, tight conditions for convergence and divergence  can be established.

\subsection{Related Work}
The  concept of signed social networks was introduced by~\cite{heider46}. His objective was to formally distinguish between friendly (positive) and hostile (negative) relationships. The notion of structural balance was introduced to understand local interactions, and formalize intricate local scenarios (e.g., {\it two of my friends are enemies}). A number of classical results on social balance was established by  \cite{harary53,harary56,davis63,davis67}, who derived critical conditions on the global structure of the social network which ensure structural balance. Social balance theory has since  become an important topic in the study of social networks. On one hand, efforts are made to characterize and compute  the degree of balance for  real-world large social networks, e.g. \cite{altafinipnas}. On the other hand,  dynamical models  are proposed for the signs of social links with the aim of describing  stable  equilibria or establishing asymptotic convergence for the sign patterns,  e.g.,  \cite{Galam1996} (where a signed structure was introduced as a revised Ising model of political coalitions, where two competing  world coalitions were  shown to have one unique stable formation), \cite{Macy2003} (who verified convergence to structural balances numerically for a Hopfield model), and~\cite{continuouspnas} (where a continuous-time dynamical model for the link signs was proposed under which convergence to structural balance was proven).

Opinion dynamics is another long-standing topic in the study of social networks, see \cite{jacksonbook} and \cite{ekbook} for recent textbooks. Following the survey~\cite{survey}, we  classify opinion evolution  models into  Bayesian and non-Bayesian updating rules. Their main difference lies in whether each node has access to and acts according to a global model or not. We refer to \cite{baysian1,baysian2} and, more recent work~\cite{acemonglubayes} for Bayesian opinion dynamics. In non-Bayesian models, nodes follow simple local updating strategies. DeGroot's model (\cite{degroot}) is a classical non-Bayesian model of opinion dynamics, where each node updates its belief as a convex combination of its neighbors' beliefs, e.g., \cite{bias,naive,Jad12}. Note that DeGroot's model  relates to averaging consensus algorithms,  e.g.,  \cite{tsithesis, xiao, boyd, jad08, fzjsac, touri, barasmc}. Non-consensus asymptotic behaviors, e.g., clustering, disagreement, and polarization, have been investigated for linear or nonlinear  variations of  DeGroot-type update rules, \cite{Krause, julien09, julien10, PNAS-biased, shi13, JSAC-Karuse}. Various models from statistical physics have also been applied to study social opinion dynamics, please refer to~\cite{StatasticPhy} for a survey.

The influence of misbehaving nodes in social networks  have been studied only to some extent.  For instance, in \cite{misinfo}, a model of the
spread of misinformation in large societies was discussed. There, some individuals are {\it forceful}, meaning that they influence the beliefs of some of the other individuals they meet, but do not change their own opinions. In \cite{como}, the authors studied the propagation of  opinion disagreement under DeGroot's model, when some nodes stick to their initial beliefs during the entire evolution. This idea was extended to binary opinion dynamics under the voter model in \cite{stubborn}.  In \cite{altafini1,altafini2}, the author proposed  a linear model for belief dynamics over signed graphs. In \cite{altafini2}, it was shown that a bipartite agreement, i.e., clustering of opinions, is reached as long as the signed social graph is strongly balanced in the sense of  the classical structural balance theory (\cite{harary56}), which presents an important  link between opinion dynamics and structure balance. However, in the model studied in~\cite{altafini1,altafini2}, all beliefs  converge to a common value, equal to zero, if the graph is not strongly balanced. This behavior  seems to be  difficult to interpret and justify from real-world observations. A game-theoretical approach   for studying the interplay between good and bad players in  collaborative networks was introduced in \cite{barasgame}.

\subsection{Contribution}
We propose and analyze a new  model for belief dynamics over signed social networks. Nodes randomly execute pairwise interactions to update their beliefs. In case of a positive link (representing that the two interacting nodes are friends), the update follows DeGroot's update rule which drives the two beliefs closer to each other. On the contrary, in case of a negative link (i.e., when the two nodes are enemies), the update  increases the difference between the two beliefs. Thus, two opposite types of opinion updates are defined, and the beliefs are driven not only by random node interactions but also by the type of relationship of the interacting nodes. Under this simple attraction--repulsion model for opinions on signed social networks, we establish a number of fundamental results on belief convergence and divergence,  and study the impact of the parameters of the update rules and of the network structure on the belief dynamics.

Using classical spectral methods, we derive conditions for mean and mean-square convergence and  divergence of beliefs. We establish phase transition phenomena for these notions of convergence, and study how the thresholds depend on the parameters of the opinion update model and on the structure of the underlying graph. We derive phase transition conditions for almost sure convergence and divergence of beliefs. The proofs are based on  what we call the {\it Triangle lemma}, which characterizes the evolution of the beliefs held by three different nodes. We utilize   probabilistic tools such as the Borel-Cantelli lemma, the Martingale convergence theorems, the strong law of large numbers, and sample-path arguments.

We  establish two counter-intuitive results about the way beliefs evolve: (i) a {\it no-survivor} theorem which states that the difference between  opinions of any two nodes tends to infinity almost surely (along a subsequence of instants)  whenever the difference between the maximum and the minimum beliefs in the network tends to infinity (along a subsequence of instants); (ii) a  {\it live-or-die} lemma which demonstrates that almost surely, the opinions either converge to an agreement or diverge. We also show that networks whose positive component includes an hypercube are (essentially, the only) robust networks in the sense that almost sure convergence of beliefs holds irrespective of the number of negative links, their positions in the network, and the strength of the negative update.

The considered model is extended to cases where updates may be asymmetric (in the sense that when two nodes interact, only one of them  updates its belief), and where beliefs have hard lower and upper constraints. The latter  boundedness  constraint adds  slight nonlinearity to the belief evolution. It turns out in this case that the classical social network structural balance theory  plays a fundamental role in determining  the asymptotic   formation of opinions:
\begin{itemize}
\item If the social network is  structurally balanced (strongly balanced, or complete and weakly balanced), i.e.,  the network can be  divided into subgroups with positive links inside each subgroup and negative links among different subgroups, then  almost surely, the  beliefs within the same subgroup will be clustered to one of the belief boundaries, when the strength of the negative updates is sufficiently large.

    \item In the absence of structural balance, and if the positive graph of the social network is connected, then almost surely,  the belief of each node oscillates between the lower and upper bounds and touches the two belief boundaries  an infinite number of times.
\end{itemize}
For balanced social networks, the boundary clustering results are established based on the almost sure happening of  suitable {\it separation} events, i.e., the node beliefs for a subgroup become group polarized (either larger or smaller than the remaining nodes' beliefs). From this argument  such events tend to happen more easily in the presence of small subgroups.   As a result,     small subgroups contribute to faster  clustering of the social  beliefs, which is consistent with the study of  minority influence in social psychology \cite{Minority1,Minority2} suggesting   that  consistent minorities can substantially  influence  opinions. For unbalanced social networks, the established opinion oscillation contributes to a new type of  belief formation which complements polarization, disagreement, and consensus \cite{PNAS-biased}.

\subsection{Paper Organization}
In Section~2, we present the signed social network model, specify the dynamics along  positive and negative links, and define the problem of interest.   Section~3 focuses on the mean and mean-square convergence  and divergence analysis, and  Section~4 considers convergence and divergence in the almost sure sense. In Section~5, we study a  model  with  upper and lower belief bounds and asymmetric updates. It is shown how structural balance determines the clustering of opinions. Finally  concluding remarks are given in Section~6.

\subsection*{Notation and Terminology}

An undirected graph is denoted by $\mathsf {G} =(\mathsf{V}, \mathsf{E} )$. Here $\mathsf{V}=\{1,\dots,n\}$  is a finite set of  vertices (nodes). Each element in $\mathsf{E}$ is an unordered pair of two distinct  nodes in $\mathsf {V}$, called an edge. The edge between nodes $i,j\in\mathsf{V}$ is denoted by $\{i,j\}$. Let $\mathsf{V}_\ast \subseteq \mathsf{V}$ be a subset of nodes. The {\it induced} graph of $\mathsf{V}_\ast$ on $\mathsf {G}$, denoted $\mathsf {G}_{\mathsf{V}_\ast}$, is the graph  $(\mathsf{V}_\ast,\mathsf{E}_{\mathsf{V}_\ast})$ with $\{u,v\}\in\mathsf{E}_{\mathsf{V}_\ast}$, $u,v\in \mathsf{V}_\ast$ if and only if $\{u,v\}\in\mathsf{E}$. A path in $\mathsf {G}$ with length $k$ is a  sequence of distinct nodes, $v_1v_2\dots v_{k+1}$,
such that  $\{v_m, v_{m+1}\} \in \mathsf{E}$, $m=1,\dots,k$. The length of a shortest path between two nodes $i$ and $j$ is called the distance between the  nodes, denoted $d(i,j)$. The  greatest length of all shortest paths is called the diameter of the graph, denoted ${\rm diam}(\mathsf{G})$. The {\it degree} matrix of $\mathsf{G}$, denoted ${\rm D}(\mathsf{G})$, is the diagonal matrix ${\rm diag}(d_1,\dots,d_n)$ with $d_i$ denoting the number of nodes sharing an edge with $i,i\in\mathsf{V}$. The {\it adjacency} matrix ${\rm A}(\mathsf{G})$ is the symmetric $n\times n$ matrix  such that $[{\rm A}(\mathsf{G})]_{ij}=1$ if $\{i,j\}\in \mathsf{E}$ and $[{\rm A}(\mathsf{G})]_{ij}=0$ otherwise.
The matrix ${\rm L}(\mathsf{G}):={\rm D}(\mathsf{G})-{\rm A}(\mathsf{G})$ is called the {\it Laplacian}  of $\mathsf{G}$. Two graphs containing  the same number of vertices are called {\it  isomorphic} if they are identical subject to a permutation of vertex labels.

All vectors are column vectors and denoted
by lower case letters. Matrices are denoted with upper case letters. Given a matrix $M$, $M'$ denotes its transpose and $M^k$ denotes the $k$-th power of $M$ when it is a square matrix. The $ij$-entry of a matrix $M$ is denoted $[M]_{ij}$.  Given a matrix $M\in \mathds{R}^{mn}$, the vectorization of $M$, denoted by ${\rm \bf vec}(M)$, is the $mn\times 1$ column vector  $([M]_{11}, \dots,  [M]_{m1},  \dots,  [M]_{1n},\dots, [M]_{mn})'$. We have ${\rm \bf vec}(ABC)=(C'\otimes A){\rm \bf vec}(B)$ for all real matrices $A,B,C$ with $ABC$ well defined. A square  matrix $M$ is called a stochastic matrix if all of its entries  are non-negative and the sum of each row of $M$ equals one. A stochastic matrix $M$ is doubly stochastic if $M'$ is also a stochastic matrix.    With the universal set prescribed,  the complement of a given set $S$ is denoted $S^c$. The orthogonal complement of a subspace $S$ in a vector space is denoted $S^\bot$.     Depending on the argument, $|\cdot|$ stands for the absolute value of a real number, the Euclidean norm of a vector, and the cardinality of a set.  Similarly with argument well defined,  $\sigma(\cdot)$ represents the $\sigma$-algebra of a random variable (vector), or the spectrum of a matrix. The smallest integer no smaller than a given real number $a$ is denoted $\lceil a\rceil$.  We use $\mathbb{P}(\cdot)$ to denote the probability, $\mathbb{E}\{\cdot\}$ the expectation,  $\mathbb{V}\{\cdot\}$ the variance  of their arguments, respectively.

\section{Opinion Dynamics over Signed Social Networks}

In this section, we present our model of interaction between nodes in a signed social network, and describe the resulting dynamics of the beliefs held by each node.

\subsection{Signed Social Network and Peer Interactions}
\begin{figure}[t]
\begin{center}
\includegraphics[height=2.7in]{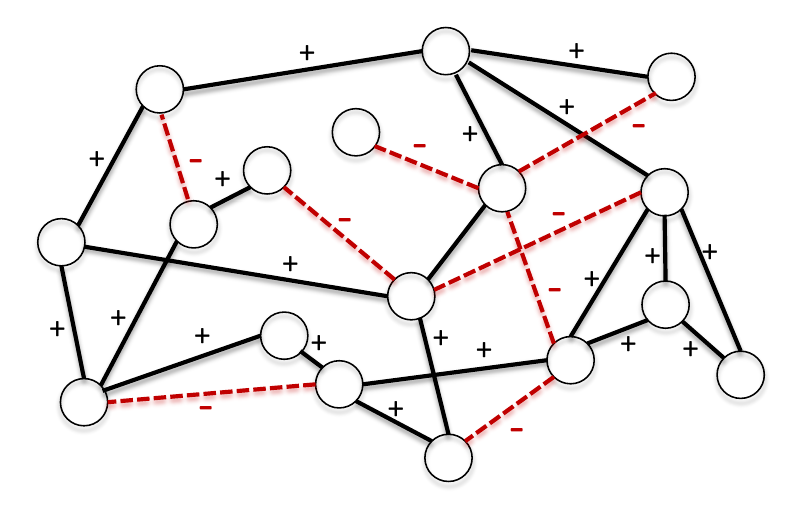}
\caption{A signed social network.} \label{frontier}
\end{center}
\end{figure}

We consider a social network with $n \geq 3$ members, each labeled by a unique integer in $\{1, 2, \dots, n\}$. The network is represented by an undirected graph $\mathsf {G} =(\mathsf{V}, \mathsf{E} )$ whose node set $\mathsf{V}=\{1, 2, \dots, n\}$ corresponds to the members and whose edge set ${\mathsf E}$ describes potential interactions between the members.  Each edge in $\mathsf{E}$ is assigned a unique label, either $+$ or $-$. In classical social network theory, a $+$ label indicates a friend relation, while a $-$ label indicates an enemy relation (\cite{heider46,harary56}).  The graph $\mathsf{G}$ equipped with a sign on each edge is then called a \emph{signed graph}. Let $\mathsf{E}_{\rm pst}$ and $\mathsf{E}_{\rm neg}$ be the collection of the positive and negative edges, respectively; clearly, $\mathsf{E}_{\rm pst} \cap\mathsf{E}_{\rm neg} =\emptyset$ and $\mathsf{E}_{\rm pst} \cup\mathsf{E}_{\rm neg} =\mathsf{E} $. We call $\mathsf{G}_{\rm pst}=(\mathsf{V}, \mathsf{E}_{\rm pst})$ and $\mathsf{G}_{\rm neg}=(\mathsf{V}, \mathsf{E}_{\rm neg})$ the positive  and the negative  graph, respectively; see Figure \ref{frontier} for an illustration. Without loss of generality, we adopt the following assumption throughout the paper.

\begin{assumption}
The underlying graph $\mathsf{G} $ is connected, and  the negative  graph $\mathsf{G}_{\rm neg}$ is nonempty.
\end{assumption}

 Actual interactions follow the model introduced in \cite{boyd}: each node initiates interactions at the instants of a rate-one Poisson process, and at each of these instants, picks a node at random to interact with. Under this model, at a given time, at most one node initiates an interaction. This allows us to order interaction events in time and to focus on modeling the node pair selection at interaction times. The node selection process is characterized by an $n\times n$ stochastic matrix $P=[p_{ij}]$ complying with the graph $\mathsf{G}$, in the sense that  $p_{ij}>0$ always implies  $\{i,j\}\in \mathsf{E}$ for $i\neq j \in \mathsf{V}$. The $p_{ij}$ represents the probability that node $i$ initiates an interaction with node $j$.  The node pair selection is then performed as follows.

\begin{definition} At each interaction event $k\geq0$, (i) a node $i\in\mathsf{V}$ is drawn uniformly at random, i.e., with probability $1/n$; (ii) node $i$ picks node $j$ with probability $p_{ij}$. In this case, we say that the unordered node pair $\{i,j\}$ is selected.
\end{definition}

The node pair selection process is assumed to be i.i.d., i.e., the nodes that initiate an interaction and the selected node pairs are identically distributed and independent over $k\ge 0$. Formally, the node selection process can be analyzed using the following probability spaces. Let $(\mathsf{E}, \mathcal{S}, \mu)$ be the probability space, where $\mathcal{S}$ is the discrete $\sigma$-algebra on $\mathsf{E}$, and $\mu$ is the probability measure defined by $\mu(\{i,j\})=\frac{p_{ij}+p_{ji}}{n}$ for all $\{i,j\}\in \mathsf{E}$. The node selection process can then be seen as a random event in the product probability space $(\Omega,\mathcal{F},\mathbb{P})$, where $\Omega=\mathsf{E}^{\mathbb{N}}=\{\omega=(\omega_0,\omega_1,\dots,): \forall k, \omega_k\in \mathsf{E}\}$, ${\cal F}={\cal S}^{\mathbb{N}}$, and $\mathbb{P}$ is the product probability measure (uniquely) defined by: for any finite subset $K\subset \mathbb{N}$, $\mathbb{P}((\omega_k)_{k\in K})=\prod_{k\in K}\mu(\omega_k)$ for any $(\omega_k)_{k\in K}\in \mathsf{E}^{|K|}$. For any $k\in \mathbb{N}$, we define the coordinate mapping $G_k:\Omega\to \mathsf{E}$ by $G_k(\omega)=\omega_k$, for all $\omega\in \Omega$ (note that $\mathbb{P}(G_k=\omega_k)=\mu(\omega_k)$), and we refer to $(G_k,k=0,1,\ldots)$ as the {\it node pair selection process}. We further refer to ${\cal F}_k=\sigma(G_0,\ldots,G_k)$ as the $\sigma$-algebra capturing the $(k+1)$ first interactions of the selection process.

\subsection{Positive and Negative  Dynamics}
Each node maintains a scalar real-valued \emph{opinion}, or \emph{belief}, which it updates whenever it interacts with other nodes. We let $x(k)\in\mathds{R}^n$ denote the vector of the beliefs held by nodes at the interaction event $k$.

The belief update depends on the relationship between the interacting nodes.  Suppose that node pair $\{i,j\}$ is selected at time $k$. The nodes that are not selected keep their beliefs unchanged, whereas the beliefs held by nodes $i$ and $j$ are updated as follows:

\begin{itemize}
\item {\it (Positive Update)} If $\{i,j\}\in\mathsf{E}_{\rm pst} $, either node $m\in\{i,j\}$  updates its belief as
\begin{align}\label{trust}
x_m(k+1)=x_m(k) + \alpha \big(x_{-m}(k)-x_m(k)\big)=(1- \alpha)x_{m}(k)+ \alpha x_{-m}(k),
\end{align}
where ${-m}\in\{i,j\}\setminus \{m\}$ and $0\leq \alpha \leq 1$.

\item {\it (Negative Update)} If $\{i,j\}\in\mathsf{E}_{\rm neg} $,  either node $m\in\{i,j\}$  updates its belief as
\begin{align}\label{130}
x_m(k+1)=x_m(k)- \beta  \big(x_{-m}(k)-x_m(k)\big)=(1+\beta)x_{m}(k)- \beta x_{-m}(k),
\end{align}
where $\beta\geq 0$.
\end{itemize}

The positive update is consistent with the classical DeGroot model (\cite{degroot}), where each node iteratively updates its belief as a convex combination of the previous beliefs of itself and of the neighbor with which it interacts. This update naturally reflects {\it trustful} or {\it cooperative} relationships. It is sometimes referred to as na\"{i}ve learning in social networks, under which {\it wisdom} can be held by the crowds (\cite{naive}). The positive update tends to drive node beliefs closer to each other and can be thought of as  the attraction of the beliefs.

The dynamics on the negative edges, on the other hand, is not yet universally agreed upon in the literature. Considerable efforts have been made to characterize these {\it mistrustful} or {\it antagonistic} relationships, which has led to a number of insightful models, e.g.,  \cite{misinfo, como, altafini1, altafini2}. Our negative update rule enforces belief differences between interacting nodes, and is the {\it opposite} of the attraction of beliefs represented by the positive update.

\subsection{Model Rationale}

\subsubsection{Relation to Non-Bayesian Rules}
Our underlying signed graph is a {\it prescribed} world with fixed trust or mistrustful relations where nodes do not switch their  relations. Two nodes holding the same opinion can be enemies, and vice versa. This contrasts Krause's model, where trustful relations are state-dependent and nodes only interact with nodes which hold similar opinions, i.e., whose beliefs are within a given distance.

In our model, the signed graph classifies the social interactions into two categories, positive and negative, each with its own type of dynamics. Studies of stubborn agents in social network~\cite{como,stubborn} also classify nodes into two categories, but stubborn agents do not account for the opinion of its neighbors. Our model is more similar to the one introduced by Altafini in~\cite{altafini2}, where the author proposed a different update rule for two nodes sharing a negative link. The model in \cite{altafini2} is written in continuous time (beliefs evolve along  some ODE), but its corresponding discrete-time update across  a negative link $\{i,j\}\in\mathsf{E}_{\rm neg}$ is:
\begin{align}\label{123}
x_m(k+1)=x_m(k)- \beta \big(x_{-m}(k)+x_m(k)\big)=(1-\beta)x_{m}(k)- \beta x_{-m}(k),\ \ m\in\{i,j\},
\end{align}
where $\beta\in (0,1)$ represents the negative strength. This update rule admits the following interpretations:
\begin{itemize}
\item  Node $i$ attempts to {\it trick} her negative neighbors $j$, by flipping  the sign of her true belief (i.e., $x_i(k)$ to $-x_i(k)$) before revealing it to $j$;
    \item  Node $i$ recognizes $j$ as her negative neighbor and upon observing  $j$'s true belief, $x_j(k)$, she tries to get  closer to the opposite view of $j$ since  $x_i(k+1)$ is a convex combination of $x_i(k)$ and $-x_j(k)$.
\end{itemize}

In both of the two interpretations of the Altafini model, the belief origin must be of some particular significance  in the nodes' belief space. This is not the case for our model, where the positive/negative dynamics describe {\it choices} intended to  {\it keeping close to friends} and  {\it keeping distance from enemies}.  When nodes $i$ and $j$ perform a negative update in our model, if $x_i(k) > x_j(k)$ then $x_i(k+1) > x_i(k)$ and if $x_i(k) < x_j(k)$ then $x_i(k+1) < x_i(k)$. That is, in either case, the node's updated opinion is in a direction away from the opinion of the interacting node (i.e., nodes make an effort to ``keep distance from the enemies" and do not assign any special meaning to the belief origin).

\begin{remark}
The Altafini model \cite{altafini2} and the current work are intended for building theories to  opinion dynamics {\rm over} signed social networks.  Indeed nontrivial efforts have been made to model the dynamics {\rm of} signed social networks themselves \cite{Galam1996}, \cite{Macy2003}, \cite{continuouspnas}. It is intriguing to ask how opinions and social networks shape each other in the presence of trustful/mistrustful relations, where fundamental difficulty arises in how to properly model such couplings as well as  the challenges brought by the couplings.
\end{remark}

\subsubsection{Relation to Bayesian Rules}

Bayesian opinion dynamics assume that there is a global {\it model} of the world and individuals aim to realize asymptotic learning of the underlying world   \cite{baysian1}, \cite{baysian2}, \newline \cite{acemonglubayes}. It has been shown that DeGroot update can also serve as a naive learning approach as long as the network somehow contains no dictators \cite{naive}.

We argue here our model corresponds to the situation where  nodes naively follow the code of keeping distance with enemies and keeping close to friends, rather than having interest in some underlying world model. Our definition of  the negative dynamics becomes quite  natural if one views the DeGroot type of update as  the approach of keeping close to friends. This simple yet informative model leads to a number of nontrivial belief formations in terms of convergence or divergence for unconstrained evolution, consensus, clustering, or oscillation under boundedness constraint.

We note that it is an interesting open challenge to find a proper  model for Bayesian learning over signed social networks, since nodes must learn in the presence of {\it negative} interactions, on the one hand, and may try to prevent their enemies from asymptotic learning, on the other.

\section{Mean and Mean-square Convergence/Divergence}
Let $x(k)=(x_1(k)\dots x_n(k))', k=0,1,\dots$ be the (random) vector of beliefs at time $k$ resulting from the node interactions. The initial beliefs $x(0)$, also denoted as $x^0$, is assumed to be deterministic. In this section, we investigate the mean and mean-square evolution of the beliefs for the considered signed social network. We introduce the following definition.

\begin{definition} (i) Belief convergence  is achieved
in \emph{expectation}   if
$\lim_{k\rightarrow \infty} \mathbb{E} \big\{x_i(k)-x_j(k) \big\}=0$ for all $i$ and $j$;
  in \emph{mean square}    if $\lim_{k\rightarrow \infty} \mathbb{E} \big\{ (x_i(k)-x_j(k))^2\big\}=0$ for all $i$ and $j$.

(ii) Belief   divergence  is achieved  in \emph{expectation}   if
$\limsup_{k\rightarrow \infty}$  $\max_{i,j} \big|\mathbb{E} \big\{ x_i(k)-x_j(k) \big\}\big|=\infty$;
  in \emph{mean square} if
$\limsup_{k\rightarrow \infty}  \max_{i,j} \mathbb{E} \big\{ (x_i(k)-x_j(k))^2 \big\}=\infty$.
\end{definition}

The belief dynamics as described above can be written as:
 \begin{align}\label{1}
 x(k+1)=W(k)x(k),
 \end{align}
where $W(k),k=0,1,\dots$ are i.i.d. random matrices satisfying
\begin{align}\label{2}
\begin{aligned}
&\mathbb{P}\Big(W(k)=\mathrm{W}^+_{ij}:= I-\alpha (e_i-e_j)(e_i-e_j)'\Big)=\frac{p_{ij}+p_{ji}}{n}, \ \ \   \{i,j\}\in \mathsf{E}_{\rm pst},\\
&\mathbb{P}\Big(W(k)=\mathrm{W}^-_{ij}:=  I+\beta(e_i-e_j)(e_i-e_j)'\Big)=\frac{p_{ij}+p_{ji}}{n}, \ \ \   \{i,j\}\in \mathsf{E}_{\rm neg},
\end{aligned}
\end{align}
and $e_m=(0 \dots 0\  1\  0 \dots 0)'$ is the $n$-dimensional unit vector whose $m$-th component is $1$. In this section, we use spectral properties of the linear system (\ref{1}) to study convergence and divergence in mean and mean-square. Our results can be seen as extensions of existing convergence results on deterministic consensus algorithms,  e.g., \cite{xiao}.

\subsection{Convergence in Mean}

We first provide conditions for convergence and divergence in mean. We then exploit these conditions to establish the existence of a phase transition for convergence when the negative update parameter $\beta$  increases. These results are illustrated at the end of this subsection. For technical reasons we adopt the following assumption in this subsection.
\begin{assumption}\label{assum-diag}
There holds either (i) $p_{ii}\geq 1/2$ for all $i\in \mathsf{V}$, or (ii) $P=[p_{ij}]$ is doubly stochastic with $n\geq 4$.
\end{assumption}

Generalization to the case when Assumption \ref{assum-diag} does not hold is  essentially straightforward but under  a bit more careful treatment.
\subsubsection{Convergence/Divergence Conditions}

Denote $P^\dag=(P+P')/n$. We write $P^\dag=P_{\rm pst}^\dag+P_{\rm neg}^\dag$, where $P_{\rm pst}^\dag$ and $P_{\rm neg}^\dag$ correspond to the positive and negative graphs, respectively.  Specifically, $[P_{\rm pst}^\dag]_{ij}=[P^\dag]_{ij}$ if $\{i,j\}\in \mathsf{E}_{\rm pst}$ and $[P_{\rm pst}^\dag]_{ij}=0$ otherwise, while $[P_{\rm neg}^\dag]_{ij}=[P^\dag]_{ij}$ if $\{i,j\}\in \mathsf{E}_{\rm neg}$ and $[P_{\rm neg}^\dag]_{ij}=0$ otherwise. We further introduce the degree matrix $D_{\rm pst}^\dag={\rm diag}(d_1^+ \dots d_n^+)$ of the positive graph, where $d_i^+= \sum_{j=1, j\neq i}^n [P_{\rm pst}^\dag]_{ij}$. Similarly, the degree matrix of the negative graph is defined as $D_{\rm neg}^\dag={\rm diag}({d}_1^- \dots {d}_n^-)$ with ${d}_i^-= \sum_{j=1, j\neq i}^n [P_{\rm neg}^\dag]_{ij}$. Then $ L_{\rm pst}^\dag=D_{\rm pst}^\dag- P_{\rm pst}^\dag$ and  $ L_{\rm neg}^\dag=D_{\rm neg}^\dag- P_{\rm neg}^\dag$ represent the (weighted) Laplacian matrices of the positive graph $ \mathsf{G}_{\rm pst}$ and negative graph $\mathsf{G}_{\rm neg}$, respectively.
It  can be easily deduced from (\ref{2})  that
\begin{align}
\mathbb{E} \{ W(k)\}= I- {\alpha} L_{\rm pst}^\dag+ {\beta} L_{\rm neg}^\dag.
\end{align}
Clearly, $\mathbf{1}'\mathbb{E} \{ W(k)\}=\mathbb{E} \{ W(k)\}\mathbf{1}=\mathbf{1}$ where $\mathbf{1}=(1 \dots  1)'$ denotes the $n\times1$ vector of  all ones,  but  $\mathbb{E} \{ W(k)\}$ is not necessarily a stochastic matrix since it may contain negative entries.

Introduce $y_i(k)=x_i(k)-\sum_{s=1}^n x_s(k)/n$ and let $y(k)=(y_1(k)\dots y_n(k))'$.  Define  $U:=\mathbf{1}\mathbf{1}'/n$  and note that
 $y(k)=(I-U)x(k)$; furthermore, $(I-U)W(k)=W(k)(I-U)=W(k)-U$ for all possible realizations of $W(k)$. Hence,  the evolution of $\mathbb{E}\{ y(k)\}$ is linear:
   \begin{align}
   \mathbb{E}\{ y(k+1)\}=\mathbb{E}\{ (I-U)W(k)x(k)\}=\mathbb{E}\{ (I-U)W(k)(I-U)x(k)\}=\big(\mathbb{E}\{ W(k)\}-U\big) \mathbb{E}\{ y(k)\}. \nonumber
   \end{align}
The following elementary inequalities
   \begin{align}\label{9}
\big|\mathbb{E}\{x_i(k)-x_j(k)\}\big|\leq \big|\mathbb{E}\{y_i(k)\}\big|+\big|\mathbb{E}\{y_j(k)\}\big|, \qquad
   \big|\mathbb{E}\{y_i(k)\}\big| \leq \frac{1}{n} \mathbb{E}\sum_{s=1}^{n}|x_i(k)-x_s(k)|
   \end{align}
imply that  belief convergence in expectation is equivalent to $\lim_{k\rightarrow \infty}|\mathbb{E}\{ y(k)\}| =0$, and belief divergence is equivalent to $\limsup_{k\rightarrow \infty}|\mathbb{E}\{ y(k)\}| =\infty$. Belief convergence or divergence is hence determined by the spectral radius of $\mathbb{E} \{ W(k)\}-U$.

With  Assumption \ref{assum-diag}, there always holds that
\begin{align*}
d_i^+= \sum_{j=1, j\neq i}^n [P_{\rm pst}^\dag]_{ij} \leq \sum_{j=1, j\neq i}^n \big(p_{ij}+p_{ji}\big)/n \leq 1/2.
\end{align*}
As a result, Ger\v{s}hgorin's Circle Theorem (see, e.g., Theorem 6.1.1 in \cite{matrix}) guarantees that each eigenvalue of $I-{\alpha}L_{\rm pst}^\dag$ is nonnegative. It then follows that  each eigenvalue of $I-{\alpha}L_{\rm pst}^\dag-U$ is nonnegative since $L_{\rm pst}^\dag U=UL_{\rm pst}^\dag=0$ and the two matrices $I-{\alpha}L_{\rm pst}^\dag$ and $U$ share the same eigenvector $\mathbf{1}$ for eigenvalue one. Moreover, it is well known in algebraic graph theory that $L_{\rm pst}^\dag$ and $L_{\rm neg}^\dag$ are positive semi-definite  matrices. As a result, Weyl's inequality (see Theorem 4.3.1 in \cite{matrix}) further  ensures that each eigenvalue of $\mathbb{E} \{ W(k)\}-U$ is also nonnegative. To summarize, we have shown that:
\begin{proposition}\label{meanthm}
Let Assumption \ref{assum-diag} hold. Belief convergence  is achieved in expectation  for all initial values if $\lambda_{\rm max}\big(I-{\alpha}L_{\rm pst}^\dag+{\beta}L_{\rm neg}^\dag-U\big)<1$; belief divergence is achieved in expectation  for almost all initial values if $\lambda_{\rm max}\big(I-{\alpha}L_{\rm pst}^\dag+{\beta}L_{\rm neg}^\dag -U\big)>1$.
\end{proposition}

In the above proposition and what follows, $\lambda_{\max}(M)$ denotes the largest eigenvalue  of the real symmetric matrix $M$, and by ``almost all initial conditions," we mean that the property holds for any initial condition $y(0)$ except if $y(0)$ is perfectly orthogonal to the eigenspace of $\mathbb{E} \{ W(k)\}-U$ corresponding to its maximal eigenvalue $\lambda_{\rm max}\big(I-{\alpha}L_{\rm pst}^\dag+{\beta}L_{\rm neg}^\dag-U\big)$. Hence the set of initial conditions where the property does not hold has zero Lebesgue measure.

The Courant-Fischer Theorem (see Theorem 4.2.11 in \cite{matrix}) implies
\begin{align}\label{6}
&\lambda_{\rm max}\big(I-{\alpha}L_{\rm pst}^\dag+{\beta}L_{\rm neg}^\dag-U\big)\nonumber =\sup_{|z|=1} z' \Big( I-{\alpha} L_{\rm pst}^\dag+ {\beta} L_{\rm neg}^\dag-U\Big)z\nonumber\\
&\quad\quad=1+\sup_{|z|=1} \bigg[-{\alpha}\sum_{\{i,j\}\in \mathsf{E}_{\rm pst}} [P^\dag]_{ij} (z_i-z_j)^2+{\beta}\sum_{\{i,j\}\in \mathsf{E}_{\rm neg}} [P^\dag]_{ij} (z_i-z_j)^2-\frac{1}{n}\big(\sum_{i=1}^n z_i \big)^2 \bigg].
\end{align}
We see from (\ref{6}) that the influence of $\mathsf{G}_{\rm pst}$ and $\mathsf{G}_{\rm neg}$ on  the belief convergence/divergence in mean are {\it separated}: links in $\mathsf{E}_{\rm pst}$ contribute to belief convergence,  while links in $\mathsf{E}_{\rm neg}$  contribute to belief divergence. As will be shown later on, this separation property no longer holds for mean-square convergence, and there may be a non-trivial correlation between the influence of $\mathsf{E}_{\rm pst}$ and that of $\mathsf{E}_{\rm neg}$.

\subsubsection{Phase Transition}

Next we study the impact of update parameters $\alpha$ and $\beta$ on the convergence in expectation. Define
$
f(\alpha,\beta):=\lambda_{\rm max}\big(I-{\alpha}L_{\rm pst}^\dag+{\beta}L_{\rm neg}^\dag-U\big).
$
The function $f$ has the following properties under Assumption \ref{assum-diag}:
\begin{itemize}
\item[(i)] {\it (Convexity)} Since both $L_{\rm pst}^\dag$ and $L_{\rm neg}^\dag$ are symmetric, $f(\alpha,\beta)$ is the spectral norm of $I-{\alpha}L_{\rm pst}^\dag+{\beta}L_{\rm neg}^\dag-U$. As every matrix norm is convex, we have
    \begin{align}
    f(\gamma(\alpha_1,\beta_1)+(1-\gamma(\alpha_2,\beta_2))\leq \gamma f(\alpha_1,\beta_1)+(1-\gamma)f(\alpha_2,\beta_2)
    \end{align}
     for all $\gamma \in [0,1]$ and $\alpha_1,\alpha_2,\beta_1,\beta_2 \in \mathds{R}$. This implies that $f(\alpha,\beta)$ is convex in $(\alpha,\beta)$.

 \item[(ii)] {\it (Monotonicity)} From (\ref{6}),  $f(\alpha,\beta)$ is non-increasing  in $\alpha$ for fixed $\beta$, and non-decreasing  in $\beta$ for fixed $\alpha$. As a result, setting $\alpha=1$ provides the {\it fastest}  convergence  whenever belief convergence in expectation is achieved (for a given fixed $\beta$). Note  that when $\alpha=1$, when two nodes interact, they simply switch their beliefs.
\end{itemize}

When $\mathsf{G}_{\rm pst}$ is connected, the second smallest eigenvalue of $L_{\rm pst}^\dag$, denoted by $\lambda_2(L_{\rm pst}^\dag)$, is positive. We can readily see that $f(\alpha,0)=1-{\alpha}\lambda_2(L_{\rm pst}^\dag)<1$. From (\ref{6}), we also have  $f(\alpha,\beta)\rightarrow \infty$ as $\beta\rightarrow \infty$ provided that $\mathsf{G}_{\rm neg}$ is nonempty. Combining these observations with the monotonicity of $f$, we conclude that:

\begin{proposition}\label{phasemean} Assume that $\mathsf{G}_{\rm pst}$ is connected and let  Assumption \ref{assum-diag} hold. Then for any fixed $\alpha \in (0,1]$, there exists a threshold value $\beta_\star >0$ (that depends on $\alpha$) such that
\begin{itemize}
\item[(i)] Belief convergence in expectation  is achieved for all initial values if  $0\leq \beta<\beta_\star $;

\item[(ii)] Belief divergence  in expectation   is achieved for almost all initial values if  $ \beta>\beta_\star $.
\end{itemize}
\end{proposition}

We remark that belief divergence can only happen for almost all initial values since if the initial beliefs of all the nodes are identical, they do not evolve over time.

\subsubsection{Examples}

 An interesting question is to determine how the phase transition threshold $\beta_\star$ scales with the network size. Answering this question seems challenging. However there are networks for which we can characterize $\beta_\star$ exactly. Next we derive explicit expressions for $\beta_\star$ when $\mathsf{G}$ is a complete graph or a ring graph. These two topologies represent the most dense and  almost the most sparse structures for a connected network.

\begin{example}[Complete  Graph]\label{example1}
Let  $\mathsf{G}=\mathsf{K}_n$, where $\mathsf{K}_n$ is the complete graph with $n$ nodes, and consider the node pair selection  matrix $P=({\mathbf{1} \mathbf{1}'-I})/{(n-1)}$. Let ${\rm L}({\mathsf{K}_n})=nI- \mathbf{1} \mathbf{1}'$ be the Laplacian  of $\mathsf{K}_n$. Then ${\rm L}({\mathsf{K}_n})$ has eigenvalue $0$ with multiplicity $1$ and eigenvalue $n$ with multiplicity $n-1$.  Define ${\rm L}(\mathsf{G}_{\rm neg})$ as the standard Laplacian of $\mathsf{G}_{\rm neg}$. Observe that
\begin{align}\label{22}
I-{\alpha}L_{\rm pst}^\dag+{\beta}L_{\rm neg}^\dag-U&=I-{\alpha}(L_{\rm pst}^\dag+L_{\rm neg}^\dag)+{(\alpha+\beta)}L_{\rm neg}^\dag-U\nonumber\\
&=I- \frac{2\alpha}{n(n-1)} {\rm L}({\mathsf{K}_n})+\frac{2(\alpha+\beta)}{n(n-1)}{\rm L}(\mathsf{G}_{\rm neg}) -U.
\end{align}
Also note that ${\rm L}(\mathsf{G}_{\rm neg}) {\rm L}({\mathsf{K}_n})={\rm L}({\mathsf{K}_n}) {\rm L}(\mathsf{G}_{\rm neg})=n  {\rm L}(\mathsf{G}_{\rm neg})$. From these observations, we can then readily conclude that:
\begin{equation}\label{eq:beta}
\beta_\star= \frac{n\alpha}{\lambda_{\rm max}({\rm L}(\mathsf{G}_{\rm neg}))}-\alpha.
\end{equation}
\end{example}

\begin{example}[Erd\H{o}s-R\'enyi Negative Graph over Complete  Graph]\label{example1} Let  $\mathsf{G}=\mathsf{K}_n$ with $$P=({\mathbf{1} \mathbf{1}'-I})/{(n-1)}.
 $$ Let $\mathsf{G}_{\rm neg}$ be the   Erd\H{o}s-R\'{e}nyi random graph (\cite{er}) where for any $i,j\in \mathsf{V}$, $\{i,j\}\in \mathsf{E}_{\rm neg}$ with probability $p$ (independently of other links). Note that since $\mathsf{G}_{\rm neg}$ is a random subgraph, the function $f(\alpha,\beta)$ becomes a random variable, and we denote by $\mathbf{P}$ the probability measure related to the randomness of the graph in Erd\H{o}s-R\'{e}nyi's model. Spectral theory for random graphs suggests that  (\cite{ding})
\begin{align}\label{120}
\frac{\lambda_{\rm max}({\rm L}(\mathsf{G}_{\rm neg}))}{pn}\rightarrow 1, \hbox{ as }n\to\infty
\end{align}
in probability. Now for fixed $p$, we deduce from (\ref{eq:beta}) and (\ref{120}) that the threshold $\beta_\star$ converges, as $n$ grows large, to $\alpha/p$ in probability. Now let us fix the update parameters $\alpha$ and $\beta$, and investigate the impact of the probability $p$ on the convergence in mean.
\begin{itemize}
\item If $p< \frac{{\alpha}}{\alpha+\beta}$, we show that $\mathbf{P}[f(\alpha,\beta) <1]\to 1$, when $n\to\infty$, i.e., when the network is large, we likely achieve convergence in mean. Let $\epsilon<\frac{{\alpha}}{(\alpha+\beta)p}-1$. It follows from (\ref{120}) that
\begin{align}
\mathbf{P} (f(\alpha,\beta)<1)&=\mathbf{P} \Big(1- \frac{2\alpha}{n(n-1)} n+\frac{2(\alpha+\beta)}{n(n-1)} \lambda_{\rm max} \big({\rm L}(\mathsf{G}_{\rm neg}))<1\Big)\nonumber\\
&=\mathbf{P} \Big({(\alpha+\beta)} \lambda_{\rm max} \big({\rm L}(\mathsf{G}_{\rm neg}))<{\alpha} n\Big)\nonumber\\
&=\mathbf{P} \Big(\frac{\lambda_{\rm max} \big({\rm L}(\mathsf{G}_{\rm neg}))}{pn}<\frac{{\alpha}}{(\alpha+\beta)p}\Big)\nonumber\\
&\geq  \mathbf{P} \Big(\Big|\frac{\lambda_{\rm max} \big({\rm L}(\mathsf{G}_{\rm neg}))}{pn}-1\Big|<\epsilon\Big)\rightarrow 1,\  \hbox{ as }n\rightarrow \infty.
\end{align}
\item If $p> \frac{{\alpha}}{\alpha+\beta}$, we similarly establish that $\mathbf{P} (f(\alpha,\beta)>1)\rightarrow 1$, when $n\rightarrow \infty$, i.e., when the network is large, we observe divergence in mean with high probability.
\end{itemize}
Hence we have a sharp phase transition between convergence and divergence in mean when the proportion of negative links $p$ increases and goes above the threshold $p_\star=\alpha/(\alpha +\beta)$.
\end{example}

\begin{example}[Ring  Graph]
Denote $\mathsf{R}_n$ as the ring graph with $n$ nodes. Let ${\rm A} ({\mathsf{R}_n})$ and ${\rm L}({\mathsf{R}_n})$ be the adjacency and Laplacian matrices of $\mathsf{R}_n$, respectively. Let the underlying graph $\mathsf{G}=\mathsf{R}_n$ with only one negative link (if one has more than two negative links, it is easy to see that divergence in expectation is achieved irrespective of $\beta >0$). Take  $P ={\rm A} ({\mathsf{R}_n})/2$.  We know that  ${\rm L}({\mathsf{R}_n})$ has eigenvalues $2-2\cos(2\pi k/n)$, $0\leq k\leq n/2$. Applying Weyl's inequality we obtain $f(\alpha,\beta)\geq 1+ ({\beta-\alpha})/{n}$. We conclude that $\beta_\star <\alpha$, irrespective of $n$.
\end{example}

\subsection{Mean-square Convergence}

We now turn our attention to the analysis of the mean-square convergence and divergence. Define:
\begin{align}
\mathbb{E}\{|y(k)|^2\}&=\mathbb{E}\{ x(k)'(I-U)x(k)\}\nonumber\\
&= x(0)'\mathbb{E}\{ W(0)\dots W(k-1)(I-U)W(k-1) \dots W(0)\}x(0).
\end{align}
Again based on inequalities (\ref{9}), we see that belief convergence in mean square is equivalent to $$
\lim_{k\rightarrow \infty}\mathbb{E}\{|y(k)|^2\} =0,
$$ and belief divergence to $\limsup_{k\rightarrow \infty}\mathbb{E}\{|y(k)|^2\} =\infty$.
Define:
\begin{align}
\Phi (k)=\begin{cases}
\mathbb{E}\{ W(0)\dots W(k-1)(I-U)W(k-1) \dots W(0)\}, & k\geq 1,\\
I-U, & k=0.
\end{cases}
\end{align}
Then, $\Phi (k)$ evolves as a linear dynamical system (\cite{fzjsac})
\begin{align}\label{122}
\Phi(k)&=\mathbb{E}\big\{ W(0)\dots W(k-1)(I-U)W(k-1) \dots W(0)\big\} \nonumber\\
&=\mathbb{E}\big\{ W(0)(I-U)W(1)\dots W(k-1)(I-U)W(k-1) \dots W(1)(I-U) W(0)\big\}\nonumber\\
&=\mathbb{E} \{ (W(k)-U) \Phi (k-1) (W(k)-U)\},
\end{align}
where in the second equality we have used the facts that $(I-U)^2=I-U$ and $(I-U)W(k)=W(k)(I-U)=W(k)-U$ for all possible realizations of $W(k)$, and the third equality is due to that $W(k)$ and $W(0)$ are i.i.d. We can rewrite (\ref{122}) using an equivalent vector form:
\begin{align}
{\rm \bf vec}(\Phi (k))= \Theta {\rm \bf vec}(\Phi (k-1)),
\end{align}
where $\Theta$ is the matrix in $\mathds{R}^{n^2\times n^2}$ given by
\begin{align}
\Theta &=\mathbb{E} \{ (W(0)-U) \otimes (W(0)-U)\}\nonumber\\
&=\sum_{\{i,j\}\in \mathsf{G}_{\rm pst}} {[P^\dag]_{ij}}\Big(\big(\mathrm{W}^+_{ij}-U\big)\otimes \big(\mathrm{W}^+_{ij}-U\big)\Big)+\sum_{\{i,j\}\in \mathsf{G}_{\rm neg}} {[P^\dag]_{ij}} \Big(\big(\mathrm{W}^-_{ij}-U\big)\otimes \big(\mathrm{W}^-_{ij}-U\big)\Big).\nonumber
\end{align}
Let $S_\lambda$ be the eigenspace corresponding to an eigenvalue $\lambda$ of $\Theta$. Define $$
\lambda_\star:=\max\{\lambda\in\sigma(\Theta): {\rm \bf vec}(I-U)\notin S_\lambda^\bot \},
$$
which denotes the spectral radius of $\Theta$ restricted to the smallest invariant subspace containing ${\rm \bf vec}(I-U)$, i.e.,  $\mathrm{S}:={\rm span}\{\Theta^k{\rm \bf vec}(I-U),k=0,1,\dots \}$. Then mean-square  belief convergence/divergence is fully determined by $\lambda_\star$: convergence in mean square for all initial conditions is achieved  if $\lambda_\star<1$, and divergence for almost all initial conditions is achieved if $\lambda_\star>1$.

Observing that $\lambda\leq 1$ for every $\lambda\in\sigma (\mathrm{W}^+_{ij})$ and  $\lambda\geq 1$ for every $\lambda\in\sigma (\mathrm{W}^-_{ij})$, we can also conclude that each link in $\mathsf{E}_{\rm pst}$ contributes positively to $\lambda_{\rm max}(\Theta)$ and each link in $\mathsf{E}_{\rm neg}$ contributes negatively  to $\lambda_{\rm max}(\Theta)$. However, unlike in the case of the analysis of convergence in expectation, although $\lambda_\star$ defines a precise threshold for the  phase-transition between mean-square convergence and divergence, it is difficult to determine the influence $\mathsf{E}_{\rm pst}$ and $\mathsf{E}_{\rm neg}$ have on $\lambda_\star$. The reason is that  they are coupled in a nontrivial manner for  the invariant subspace $\mathrm{S}$.  Nevertheless, we are still able to propose the following conditions for mean-square belief convergence and divergence:

\begin{proposition}\label{meansquarethm} Belief convergence  is achieved for all initial values in mean square  if $$\lambda_{\rm max}\big(I-{2\alpha(1-\alpha)}L_{\rm pst}^\dag+{2\beta(1+\beta)}L_{\rm neg}^\dag -U\big)<1;
 $$
 belief divergence is achieved in mean square   for almost all initial values if   $$\lambda_{\rm max}\big(I-{\alpha}L_{\rm pst}^\dag+{\beta}L_{\rm neg}^\dag -U\big)>1$$ or $$\lambda_{\rm min}\big(I-{2\alpha(1-\alpha)}L_{\rm pst}^\dag+{2\beta(1+\beta)}L_{\rm neg}^\dag -U \big)>1.
 $$
\end{proposition}

The condition  $\lambda_{\rm max}\big(I-{\alpha}L_{\rm pst}^\dag+{\beta}L_{\rm neg}^\dag -U\big)$ is sufficient for  mean square divergence, in view of Proposition \ref{meanthm} and the fact that $\mathscr{L}_1$ divergence implies $\mathscr{L}_p$ divergence for all $p\geq1$. The other  conditions are essentially  consistent with  the upper and lower bounds of $\lambda_\star$ established in Proposition~4.4 of \cite{fzjsac}.  Proposition~\ref{meansquarethm} is a consequence of Lemma \ref{lem2} (see Appendix), as explained in Remark \ref{remarkproof}.

\section{Almost Sure Convergence/Divergence}

In this section, we explore the almost sure convergence of beliefs in signed social networks. We introduce the following definition.

\begin{definition} Belief convergence  is achieved \emph{almost surely} (a.s.)  if $
\mathbb{P}\big(\lim_{k\rightarrow \infty} \big| x_i(k)-x_j(k)\big|=0\big)=1
$
for all $i$ and $j$;   Belief   divergence  is achieved  {\em almost surely} if
$\mathbb{P}\big(\limsup_{k\rightarrow \infty} \max_{i,j} \big |x_i(k)-x_j(k)\big|=\infty\big)=1$.
\end{definition}

Basic probability theory tells us that  mean-square belief convergence implies belief convergence in expectation (mean convergence), and similarly belief divergence in expectation implies belief divergence in mean square. However, in general there is no direct connection between almost sure convergence/divergence and mean or mean-square convergence/divergence. Finally observe that, a priori, it is not clear that either a.s. convergence or a.s. divergence should be achieved.

While the analysis of the convergence of beliefs in mean and square-mean mainly relied on spectral arguments, we need more involved probabilistic methods (e.g., sample-path arguments, martingale convergence theorems) to study almost sure convergence or divergence. We first establish two insightful properties of the belief evolutions: (i) the {\it no-survivor} property stating that in case of almost sure divergence, the difference between the beliefs of any two nodes in the network tends to infinity  (along a subsequence of instants); (ii) the {\it live-or-die} property which essentially states that the maximum difference between the beliefs of any two nodes either grows to infinity, or vanishes to zero. We then show a zero-one law and a phase transition of almost sure convergence/divergence. Finally, we investigate the robustness of networks against negative links. More specifically, we show that when the graph $\mathsf{G}_{\rm pst}$ of positive links contains an hypercube, and when the positive updates are truly averaging, i.e., $\alpha=1/2$, then almost sure belief convergence is reached in finite time, irrespective of the number of negative links, their positions in the network, and the negative update parameter $\beta$. We believe that these are the only networks enjoying this strong robustness property.

\subsection{The No-Survivor Theorem}

The following theorem establishes that in the case of almost sure divergence, there is no pair of nodes that can  {\it survive} this divergence: for any two nodes, the difference in their beliefs  grow arbitrarily large.

\begin{theorem}\label{nosurvivor} ({\it No-Survivor})
Fix the initial condition and assume almost sure belief divergence. Then $\mathbb{P}\big(\limsup_{k\rightarrow \infty}  \big |x_i(k)-x_j(k)\big|=\infty\big)=1$ for all $i\neq j\in\mathsf{V}$.
\end{theorem}

Observe that the above result only holds for the almost sure divergence. We may easily build simple network examples where we have belief divergence in expectation (or mean square), but where some node pairs survive, in the sense that the difference in their beliefs vanishes  (or at least bounded). The no-survivor theorem indicates that to check almost sure divergence, we may just observe the evolution of beliefs held at two arbitrary nodes in the network.

\subsection{The Live-or-Die lemma and Zero-One Laws}

Next we further classify the ways beliefs can evolve. Specifically, we study the following events: \\
for any initial beliefs $x^0$,
\begin{align}
{\mathscr{C}}_{x^0}\doteq \big\{\limsup_{k\rightarrow \infty} \max_{i,j} |x_i(k)-x_j(k)|=0  \big\}, &\quad \mathscr{D}_{x^0}\doteq \big\{ \limsup_{k\rightarrow \infty} \max_{i,j} |x_i(k)-x_j(k)|=\infty   \big\}, \nonumber\\
{\mathscr{C}}_{x^0}^\ast\doteq \big\{\liminf_{k\rightarrow \infty} \max_{i,j} |x_i(k)-x_j(k)|=0  \big\}, &\quad \mathscr{D}_{x^0}^\ast\doteq \big\{ \liminf_{k\rightarrow \infty} \max_{i,j} |x_i(k)-x_j(k)|=\infty   \big\},\nonumber
\end{align}
and
\begin{align*}
\mathscr{C}\doteq &\big\{\limsup_{k\rightarrow \infty} \max_{i,j}|x_i(k)-x_j(k)|=0 \mbox{ for all $x^0\in\mathds{R}^n$}  \big\},\\
 \mathscr{D}\doteq &\big\{ \exists\ \mbox{(deterministic)}\ x^0\in\mathds{R}^n,\  \mbox{ s.t.}\  \limsup_{k\rightarrow \infty} \max_{i,j} |x_i(k)-x_j(k)|=\infty   \big\}.
\end{align*}

We establish that the maximum difference between the beliefs of any two nodes either goes to $\infty$, or to zero. This result is referred to as {\it live-or-die} lemma:

\begin{lemma}\label{thmliveordie}  ({\it  Live-or-Die})
Let $\alpha\in (0,1)$ and $\beta>0$.  Suppose $\mathsf{G}_{\rm pst}$ is connected. Then (i) $\mathbb{P}(\mathscr{C}_{x^0})+\mathbb{P}(\mathscr{D}_{x^0})=1$; (ii)  $\mathbb{P}(\mathscr{C}_{x^0}^\ast)+\mathbb{P}(\mathscr{D}_{x^0}^\ast)=1$.  \\
As a consequence,  almost surely, one of the following events happens:
\begin{align*}
&\big\{\lim_{k\rightarrow \infty} \max_{i,j} |x_i(k)-x_j(k)|=0\big\}; \nonumber\\
&\big\{\lim_{k\rightarrow \infty} \max_{i,j} |x_i(k)-x_j(k)|=\infty\big\}; \nonumber\\
&\big\{ \liminf_{k\rightarrow \infty} \max_{i,j} |x_i(k)-x_j(k)|=0;\  \limsup_{k\rightarrow \infty} \max_{i,j} |x_i(k)-x_j(k)|=\infty\big\}. \nonumber
\end{align*}
\end{lemma}

The Live-or-Die lemma deals with events where the initial beliefs have been fixed. We may prove stronger results on the probabilities of events that hold for {\it any} initial  condition, e.g., $\mathscr{C}$, or for at least one initial condition, e.g., $\mathscr{D}$:

\begin{theorem}\label{thmzeroone} ({\it  Zero-One Law})
Let $\alpha\in[0,1]$ and $\beta>0$. Both $\mathscr{C}$ and $\mathscr{D}$ are trivial events (i.e., each of them occurs with probability equal to either 1 or 0)  and $\mathbb{P}(\mathscr{C})+\mathbb{P}(\mathscr{D})=1$.
\end{theorem}

To prove this result, we show that $\mathscr{C}$ is a tail event, and hence trivial in view of Kolmogorov's zero-one law (the same kind of arguments has been used by \cite{jad08}). From the Live-or-Die lemma, we then simply deduce that $\mathscr{D}$ is also a trivial event. Note that  $\mathscr{C}_{x^0}$ and $\mathscr{D}_{x^0}$ may not be trivial events. In fact, we can build examples where $\mathbb{P}(\mathscr{C}_{x^0})=1/2$ and $\mathbb{P}(\mathscr{D}_{x^0})=1/2$.

\subsection{Phase Transition}

As for the convergence in expectation, for fixed positive update parameter $\alpha$, we are able to establish the existence of thresholds for the value $\beta$ of the negative update parameter, which characterizes  the almost sure belief convergence and divergence.

\begin{theorem}\label{thmphasetran} ({\it  Phase Transition})
Suppose  $\mathsf{G}_{\rm pst}$ is connected. Fix $\alpha\in (0, 1) $ with $\alpha\neq 1/2$. Then

(i) there exists $\beta^\natural(\alpha) >0$ such that
$\mathbb{P}(\mathscr{C})=1$ if  $0\leq \beta<\beta^\natural $;

(ii) there exists $\beta^\sharp(\alpha) >0$ such that  $\mathbb{P}(\liminf_{k\rightarrow \infty} \max_{i,j}|x_i(k)-x_j(k)|=\infty)=1$ for almost all initial values  if  $ \beta>\beta^\sharp$.
\end{theorem}

It should be observed that the divergence condition in (ii) is stronger than our notion of almost sure belief divergence (the maximum belief difference between two nodes diverge almost surely to $\infty$). Also note that $\beta^\natural \le \beta^\sharp$, and we were not able to show that the gap between these two thresholds vanishes (as in the case of belief convergence in expectation or mean-square).

\subsection{Robustness to Negative Links: the Hypercube}

We have seen in Theorem \ref{thmphasetran} that when $\alpha\neq 1/2$, one single negative link is capable of driving the network beliefs to almost sure divergence as long as $\beta$ is sufficiently large. The following result shows that  the evolution of the beliefs can be  robust against negative links. This is the case when nodes can reach an agreement in finite time. In what follows, we provide conditions on $\alpha$ and the structure of the graph under which finite time belief convergence is reached.

\begin{proposition}\label{propcubic}
Suppose  there exist an integer $T\geq1$ and a finite sequence of node pairs $\{i_s,j_s\}\in \mathsf{G}_{\rm pst},s=1,2,\dots, T$   such that
$\mathrm{W}^+_{i_Tj_T}\cdots \mathrm{W}^+_{i_1j_1}=U$. Then   $\mathbb{P}(\mathscr{C})=1$ for all  $ \beta\geq 0$.
\end{proposition}

Proposition \ref{propcubic} is a direct consequence of the Borel-Cantelli Lemma.  If there is  a finite sequence of node pairs $\{i_s,j_s\}\in \mathsf{G}_{\rm pst},s=1,2,\dots, T$   such that
$\mathrm{W}^+_{i_Tj_T}\cdots \mathrm{W}^+_{i_1j_1}=U$, then
$$
\mathbb{P}\Big( W({k+T})\cdots W({k+1})=U\Big)\geq \Big(\frac{p_\ast}{n}\Big)^{T},
$$
for all $k\geq 0$, where $p_\ast=\min\{ p_{ij}+p_{ji}:\{i,j\}\in\mathsf{E}\}$.
Noting that $UW(k)=W(k)U=U$ for all possible realizations of $W(k)$,  the Borel-Cantelli Lemma guarantees that   $$
\mathbb{P}\Big( \lim_{k\rightarrow \infty} W(k)\cdots W(0)=U\Big)=1
$$
for all $\beta\geq 0$, or equivalently,  $\mathbb{P}(\mathscr{C})=1$ for all $\beta\geq 0$. This proves Proposition \ref{propcubic}.

The existence of such finite sequence of node pairs under which the beliefs of the nodes in the network reach a common value in finite time is crucial (we believe that this condition is actually necessary) to ensure that the influence of $\mathsf{G}_{\rm neg}$ vanishes. It seems challenging to know whether this is at all possible. As it turns out, the structure of the positive graph plays a fundamental role. To see that, we first provide some definitions.

\begin{figure}[t]
\begin{center}
\includegraphics[height=1.6in]{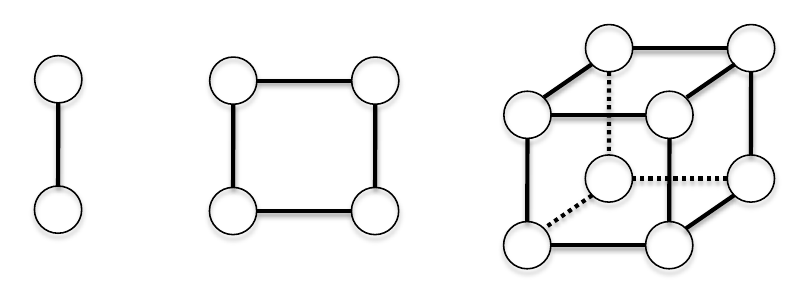}
\caption{The hypercubes $\mathsf{H}^1$, $\mathsf{H}^2$, and $\mathsf{H}^3$.} \label{hypercube}
\end{center}
\end{figure}

\begin{definition}
Let $\mathsf{G}_1=(\mathsf{V}_1,\mathsf{E}_1)$ and $\mathsf{G}_2=(\mathsf{V}_2,\mathsf{E}_2)$ be a pair of graphs. The Cartesian product of  $\mathsf{G}_1$ and $\mathsf{G}_2$, denoted by $\mathsf{G}_1\square\mathsf{G}_2$, is defined by

 (i) the  vertex set of $\mathsf{G}_1\square\mathsf{G}_2$ is $\mathsf{V}_1 \times \mathsf{V}_2$, where  $\mathsf{V}_1 \times \mathsf{V}_2$ is the Cartesian product of $\mathsf{V}_1$ and $\mathsf{V}_2$;

  (ii) for any  two vertices $(v_1,v_2), (u_1,u_2)\in \mathsf{V}_1 \times \mathsf{V}_2$, there is an edge between them in $\mathsf{G}_1\square\mathsf{G}_2$ if and only if either $v_1=u_1$ and $\{v_2,u_2\}\in \mathsf{E}_2$, or $v_2=u_2$ and $\{v_1,u_1\}\in \mathsf{E}_1$.

Let $\mathsf{K}_2$ be the complete graph with two nodes. The  $m$-dimensional Hypercube $\mathsf{H}^{m}$ is then  defined as
\begin{align}
\mathsf{H}^{m}=\underbrace{\mathsf{K}_2\square\mathsf{K}_2 \dots \square\mathsf{K}_2}_{m\  {\rm times}} \nonumber.
\end{align}
\end{definition}
An illustration of hypercubes is in Figure \ref{hypercube}.

The following result provides sufficient conditions to achieve finite-time convergence.
\begin{proposition}\label{quasicub1}
If $\alpha=1/2$, $n=2^m$ for some integer $m>0$, and $\mathsf{G}_{\rm pst}$ has a subgraph  isomorphic with an $m$-dimensional  hypercube, then there exists a sequence of $(n \log_2 n)/2$ node pairs $\{i_s,j_s\}\in \mathsf{G}_{\rm pst},s=1,\dots, (n \log_2 n)/2$   such that
$\mathrm{W}^+_{i_{(n \log_2 n)/2}j_{(n \log_2 n)/2}}\cdots \mathrm{W}^+_{i_1j_1}=U$.
\end{proposition}

Next we derive necessary conditions for finite time convergence. Let us first recall the following definition.

\begin{definition}
Let $\mathsf{G}=(\mathsf{V},\mathsf{E})$ be a graph. A {\it matching} of $\mathsf{G}$ is a set of pairwise non-adjacent edges in the sense that  no two edges share a common vertex.  A {\it perfect  matching} of $\mathsf{G}$ is a matching which matches all vertices.
\end{definition}

\begin{proposition}\label{quasicub2}
 If  there exist an integer $T\geq 1$ and a  sequence of  node pairs $\{i_s,j_s\}\in \mathsf{G}_{\rm pst},s=1,2,\dots, T$ such that $\mathrm{W}^+_{i_{T}j_{T}}\cdots \mathrm{W}^+_{i_1j_1}=U$, then $\alpha=1/2$, $n=2^m$, and $\mathsf{G}_{\rm pst}$ has a perfect matching.
\end{proposition}

In fact, in the proof of Proposition \ref{quasicub2}, we show that if $\mathrm{W}^+_{i_{T}j_{T}}\cdots \mathrm{W}^+_{i_1j_1}=U$, then a subset of  $$
\big\{\{i_1,j_1\}, \dots, \{i_{T},j_{T}\}\big\}
$$ forms  a perfect matching of $\mathsf{G}_{\rm pst}$.

We have seen that the belief dynamics and convergence can be robust against negative links, but this robustness comes at the expense of strong conditions on the number of the nodes and  the structure of the positive graph.

\section{Belief Clustering and Structural Balance}

So far we have studied the belief dynamics when the node interactions are symmetric, and the values of beliefs are unconstrained. The results illustrate that often either convergence or divergence can be predicted for the social-network beliefs. Although this symmetric and unconstrained belief update rule is plausible for ideal social network models, in reality these assumptions might  not hold, that is:  when $\{i,j\}$ is selected, it might happen that only one of the two nodes in $i$ and $j$ updates its belief;
 there might be  a hard constraint on beliefs: $x_i(k)\in [-A,A]$  for all $i$ and $k$, and for some $A>0$.

In this section, we consider the following model for the updates of the beliefs. Define:
\begin{align}\mathscr{P}_A(z)=
\begin{cases}
-A, & \mbox{if}\ z< -A;\\
z, & \mbox{if}\ z\in [-A,A];\\
A, & \mbox{if}\ z>A.
\end{cases}
\end{align}
Let $a,b,c>0$ be three positive real numbers such that $a+b+c=1$, and define the function $\theta:\mathsf{E}\to \mathbb{R}$ so that $\theta(\{i,j\})=\alpha$ if $\{i,j\}\in\mathsf{E}_{\rm pst} $ and $\theta(\{i,j\})=-\beta$ if  $\{i,j\}\in\mathsf{E}_{\rm neg}$.
Assume that node $i$ interacts with node $j$ at time $k$. Nodes $i$ and $j$ update their beliefs as:

\medskip

\noindent {\it [Asymmetric and Constrained Belief Evolution]}
\begin{align}\label{121}
\begin{aligned}
 &x_i(k+1)=\mathscr{P}_A\big((1-\theta)x_{i}(k)+ \theta x_{j}(k)\big)\ \mbox{and}\ x_{j}(k+1)=x_j(k),\ & \mbox{with probability $a$}; \\
 &x_j(k+1)=\mathscr{P}_A\big((1-\theta)x_{j}(k)+ \theta x_{i}(k)\big)\ \mbox{and}\  x_{i}(k+1)=x_i(k),\ & \mbox{with probability $b$}; \\
 &x_m(k+1)=\mathscr{P}_A\big((1-\theta)x_{m}(k)+ \theta x_{-m}(k)\big), \ m\in\{i,j\}, \ & \mbox{with probability $c$}.
\end{aligned}
\end{align}

\medskip

Enforcing the belief within the  interval $[-A,A]$ can be viewed as a social member's decision based on her fundamental model of the world.   With asymmetric and constrained belief evolution, the  dynamics become essentially nonlinear, which brings new challenges in the analysis. We continue to use $\mathbb{P}$ to denote the overall probability measure capturing the randomness of the updates in the asymmetric constrained model.


\subsection{Balanced Graphs and Clustering}

We introduce the notion of {\it balance} for signed  graphs, for which we refer to \cite{wasserman} for a comprehensive  discussion.

\begin{definition}
Let $\mathsf{G}={(\mathsf{V},\mathsf{E})}$ be a signed graph. Then

(i) $\mathsf{G}$ is {\it weakly balanced} if there is an integer $k\geq2$ and a partition of $\mathsf{V}=\mathsf{V}_1\cup \mathsf{V}_2 \dots \cup\mathsf{V}_k$, where $\mathsf{V}_1,\dots,\mathsf{V}_k$ are nonempty and  mutually disjoint,  such that any edge between different $\mathsf{V}_i$'s is negative, and any edge within each $\mathsf{V}_i$ is positive.

(ii) $\mathsf{G}$ is {\it strongly balanced} if it is weakly balanced with $k=2$.
\end{definition}

Harary's balance theorem states that a signed graph $\mathsf{G}$ is strongly balanced if and only if there is no cycle with an odd number of negative edges in $\mathsf{G}$ (\cite{harary56}), while $\mathsf{G}$ is weakly  balanced if and only if no cycle has exactly one negative edge in $\mathsf{G}$ (\cite{davis67}).

It turned out that, with certain balance of the underlying graph, clustering arises for the social-network beliefs. We make the following definition.
\begin{definition} (i) Let $\mathsf{G}$ be strongly balanced subject to partition $\mathsf{V}=\mathsf{V}_1\cup \mathsf{V}_2$. Then almost sure
boundary belief clustering for the initial value  $x^0$ is achieved  if there are two random variables $B_1^\dag(x^0)$ and $B_2^\dag(x^0)$, both taking values in $\{-A,A\}$, such that:
\begin{align}
\mathbb{P}\Big( \lim_{k\rightarrow \infty} x_i(k)=B_1^\dag(x^0),  i\in \mathsf{V}_1;\  \lim_{k\rightarrow \infty} x_i(k)=B_2^\dag(x^0),  i\in \mathsf{V}_2\Big)=1.
\end{align}

(ii) Let $\mathsf{G}$ be weakly  balanced subject to partition $\mathsf{V}=\mathsf{V}_1\cup \mathsf{V}_2 \dots \cup\mathsf{V}_m$ for some $m\geq 2$. Then almost sure
boundary belief  clustering for the initial value  $x^0$ is achieved if there are  there are $m$ random variables, $B_1^\sharp(x^0),\dots,B_m^\sharp(x^0)$, each of which  taking values in $\{-A,A\}$, such that:
\begin{align}
\mathbb{P}\Big( \lim_{k\rightarrow \infty} x_i(k)=B_j^\sharp(x^0),\   i\in \mathsf{V}_j,\  j=1,\dots,m\Big)=1.\
\end{align}
\end{definition}

In the case of strongly balanced graphs, we can show that beliefs are asymptotically clustered when $\beta$ is large enough, as stated in the following theorem.

\begin{theorem}\label{strongbalance}
Assume that $\mathsf{G}$ is strongly balanced under partition $\mathsf{V}=\mathsf{V}_1\cup \mathsf{V}_2$, and that $\mathsf{G}_{\mathsf{V}_1}$ and $\mathsf{G}_{\mathsf{V}_2}$ are connected. For any $\alpha\in (0,1)\setminus\{ 1/2\}$, when $\beta$ is sufficiently large, for almost all initial values $x^0$, almost sure boundary  belief clustering is achieved under the update rule (\ref{121}).
\end{theorem}

In fact, there holds  $B_1^\dag(x^0)+B_2^\dag(x^0)=0$  almost surely in the above boundary belief clustering for strongly balanced social networks. Theorem \ref{strongbalance} states that, for strongly  balanced social networks, beliefs are eventually polarized to the two opinion boundaries.

The analysis of belief dynamics in weakly balanced graphs is more involved, and we restrict our attention to complete graphs. In social networks, this case means that everyone knows everyone else -- which constitutes a suitable model for certain social groups of small sizes (a classroom, a sport team, or the UN, see \cite{ekbook}). As stated in the following theorem, for  weakly balanced complete  graphs, beliefs are again clustered.

\begin{theorem}\label{thmweakbalance}
Assume that $\mathsf{G}$ is a complete and weakly balanced graph under the partition $\mathsf{V}=\mathsf{V}_1\cup  \mathsf{V}_2 \dots \cup\mathsf{V}_m$ with $m\geq 2$. Further assume that $\mathsf{G}_{\mathsf{V}_j}, j=1,\dots,m$ are connected. For any $\alpha\in (0,1)\setminus\{ 1/2\}$, when $\beta$ is sufficiently large,  almost sure boundary  belief clustering is achieved for almost all initial values under (\ref{121}).
\end{theorem}

\begin{remark}
Under the model (\ref{123}),  it can be shown (cf., \cite{altafini2}, \cite{TCNS})
 \begin{itemize}
  \item[(i)] if $\mathsf{G}$ is strongly balanced and $\beta \in(0,1)$, then there are two values $z_1(x^0)$ and $z_2(x^0)$ such that
\begin{align}\label{altafini-convergence}
\mathbb{P}\big(\lim_{k\rightarrow \infty}x_i(k)=z_1(x^0), i\in\mathsf{V}_1,\ \ \lim_{k\rightarrow \infty}x_i(k)=z_2(x^0), i\in \mathsf{V}_2\big)=1.
\end{align}

 \item[(ii)] if $\mathsf{G}$ is not strongly balanced (i.e., even if it is weakly balanced) and $\beta \in(0,1)$, then
\begin{align}
\mathbb{P}\big(\lim_{k\rightarrow \infty}x_i(k)=0,\ i\in\mathsf{V}\big)=1,
\end{align}
where the impact of the initial beliefs are entirely erased from the asymptotic limit.
\end{itemize}

Our Theorem \ref{strongbalance} appears to be similar to (\ref{altafini-convergence}), but the clustering  in Theorem \ref{strongbalance} is due to  fundamentally different reasons: besides the strong balance of the social network, it is the nonlinearity in the constrained update ($\mathscr{P}_A(\cdot)$), and the sufficiently large $\beta$ that makes the boundary clustering arise in Theorem \ref{strongbalance}. In contrast, (\ref{altafini-convergence}) is resulted from the crucial condition that  $\beta\in (0,1)$. Under the Altafini model (\ref{123}), even when $\beta$ is sufficiently large, it is easy to see that  the boundary clustering in Theorem \ref{thmweakbalance} can never happen for weakly balanced graphs.
\end{remark}

The distribution of the clustering limits  established in Theorems \ref{strongbalance} and \ref{thmweakbalance} relies on the initial value. In this way, the initial beliefs make an impact on the final belief limit, which is either $A$ or $-A$. The boundary clustering is due to the hard boundaries of the beliefs  as well as the negative updates (ironically the larger the better), whose mechanism is fundamentally different with  the opinion clustering phenomena resulted from missing of connectivity in Krause types of models \cite{Krause,julien09,julien10,JSAC-Karuse}, or  nonlinear bias in the opinion evolution \cite{PNAS-biased}.

\begin{remark}
Note that in the considered asymmetric and constrained belief evolution, we take symmetric  belief boundaries $[-A,A]$ just for simplifying the discussion. Theorems \ref{strongbalance} and \ref{thmweakbalance} continue to hold if the belief boundaries are chosen to be $[A,B]$ for arbitrary $-\infty<A<B<\infty$ \footnote{This further confirms that, in our model,  the origin  of the belief space has no special meaning at all, in contrast to the model of  \cite{altafini2}.}. Letting $A=0, B=1$, our boundary clustering results in Theorems \ref{strongbalance} and \ref{thmweakbalance} are then formally consistent with the belief polarization result, Theorem 3,  in \cite{PNAS-biased}. It is worth mentioning that Theorem 3 in \cite{PNAS-biased} relies on a type of strong balance (the two-island assumption) and that the initial beliefs should be separated, while Theorems  \ref{strongbalance} and \ref{thmweakbalance} hold for almost arbitrary initial values.
\end{remark}

The proof of Theorems  \ref{strongbalance} and \ref{thmweakbalance}  is obtained by  establishing the almost sure happening of  suitable {\it separation} events, i.e., the node beliefs for a subnetwork become group polarized (either larger or smaller than the remaining nodes' beliefs). From the analysis it is clear that such events tend to happen more easily for small subnetworks in the partition of (strongly or weakly) balanced social networks. On the other hand,  boundary belief clustering follows quickly after  the separation event, even in the presence of  large subgroups. {For a  large subgroup, the boundary clustering to a consensus for its members is  more a consequence of the ``push" by the already separated small subgroups, rather than the trustful interactions therein}.  This means,  relatively  small subgroups contribute to faster occurrence of the clustering of the entire social network beliefs. Therefore, these results are in strong consistency with the research of minority influence in social psychology \cite{Minority1,Minority2},   which suggests  that  consistent
minorities can substantially  modify people's private attitudes and opinions.

\subsection{When Balance is Missing}

Since the boundary constraint only restricts the negative update, similar to Theorem \ref{thmphasetran},  for sufficiently small $\beta$, almost sure state consensus can be guaranteed when the positive graph $\mathsf{G}_{\rm pst}$ is connected.

In absence of any balance property for the underlying graph, belief clustering may not happen. However, we can establish that when the positive graph is connected, then clustering cannot be achieved when $\beta$ is large enough. In fact, the belief of a given node touches the two boundaries $-A$ and $A$ an infinite number of times. Note that if the positive graph is connected, then the graph cannot be balanced.

\begin{theorem}\label{thmergodic}
Assume that the positive graph $\mathsf{G}_{\rm pst}$ is connected. For any $\alpha\in (0,1)\setminus\{ 1/2\}$, when $\beta$ is sufficiently large, for almost all initial beliefs, under (\ref{121}), we have: for all $i\in \mathsf{V}$,
\begin{align}
\mathbb{P}\Big( \liminf_{k\rightarrow \infty} x_i(k)=-A,\  \limsup_{k\rightarrow \infty} x_i(k)=A\Big)=1.
\end{align}
\end{theorem}

Theorem \ref{thmergodic} suggests  a new class of collective formation for the social beliefs beyond consensus, disagreement, or clustering studied in the literature.

\begin{remark}
The condition that $\beta$ being sufficiently large in Theorems \ref{strongbalance} and \ref{thmweakbalance} is just a technical assumption ensuring  almost sure boundary clustering. Practically one can often encounter such clustering even for a small $\beta$, as illustrated in the coming numerical examples. On the other hand, Theorem \ref{thmergodic} relies crucially on a large $\beta$, while a small $\beta$ leads to belief consensus even in the presence of the negative edges.
\end{remark}

\subsection{Numerical Examples}
We now provide a few numerical examples to illustrate the results established in this section. We take $A=1$ so that the node beliefs are restricted to the interval $[-1,1]$. We take $\alpha=1/3$ for the positive dynamics and $a=b=c=1/3$ for the random asymmetric updates. The pair selection process is given by that when a node $i$ is drawn, it will choose one of its neighbors with equal probability $1/{\rm deg}(i)$, where ${\rm deg}(i)$ is the degree of node $i$ in the underlying graph $\mathsf{G}$.

\begin{figure}
\begin{minipage}[t]{0.5\linewidth}
\centering
\includegraphics[width=2.8in]{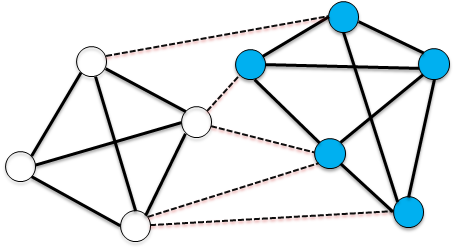}
\end{minipage}%
\begin{minipage}[t]{0.5\linewidth}
\centering
\includegraphics[width=2.8in]{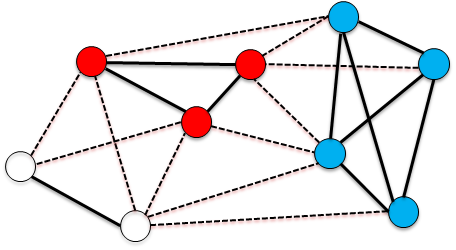}
\end{minipage}
\caption{Strongly balanced (left) and  weakly balanced (right) social graphs. The negative links are shadowed. Nodes within the same subgraph in the balance partition are marked with the same color.  }
\label{fig:examplegraphs}
\end{figure}

\begin{figure}
\begin{minipage}[t]{0.5\linewidth}
\centering
\includegraphics[width=2.8in]{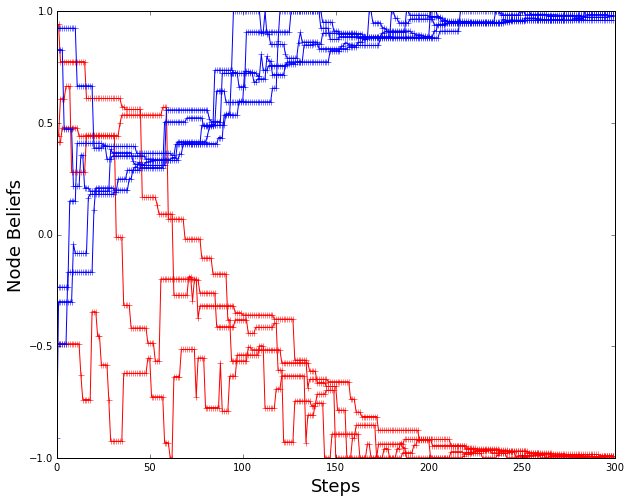}
\end{minipage}%
\begin{minipage}[t]{0.5\linewidth}
\centering
\includegraphics[width=2.8in]{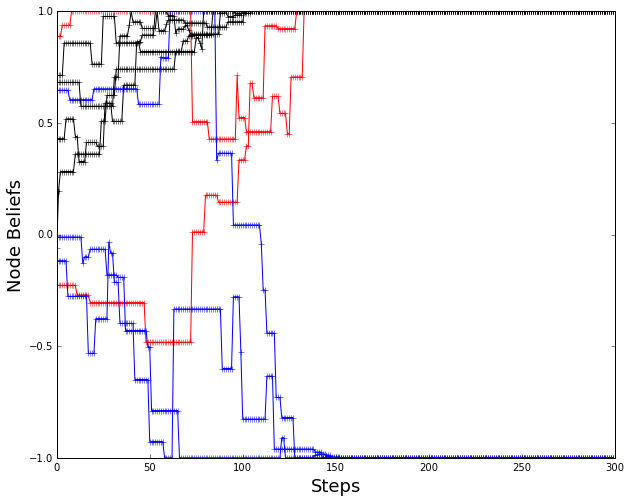}
\end{minipage}
\caption{The evolution of beliefs for strongly balanced (left) and  weakly balanced (right)  graphs. The beliefs of nodes within the same subgraph in the balance partition are marked with the same color.}
\label{fig:balanceEvolution}
\end{figure}

First of all we select two social graphs, one is strongly balanced and the other is weakly balanced, as shown in Figure \ref{fig:examplegraphs}. We take $\beta=0.2$ and randomly select the nodes' initial values. It is observed that the boundary clustering phenomena established  in Theorems \ref{strongbalance} and \ref{thmweakbalance} practically show up in every run of the random belief updates. We plot one of their typical sample paths in Figure \ref{fig:balanceEvolution}, respectively, for the strongly balanced and weakly balanced graphs in Figure \ref{fig:examplegraphs}. In fact one can see that the clustering is achieved in around $300$ steps.

\begin{figure}[t]
\begin{center}
\includegraphics[width=3.6in]{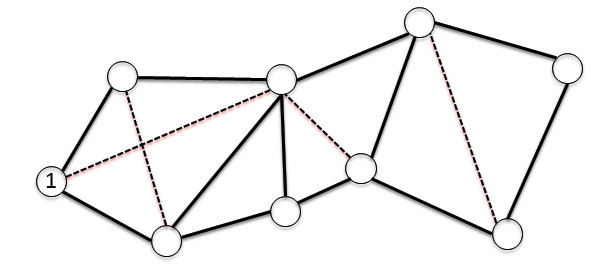}
\caption{A social network which is neither strongly nor weakly balanced. The negative links are dashed.}
\label{fig:notbalancegraph}
\end{center}
\end{figure}

\begin{figure}[t]
\begin{center}
\includegraphics[width=4.8in]{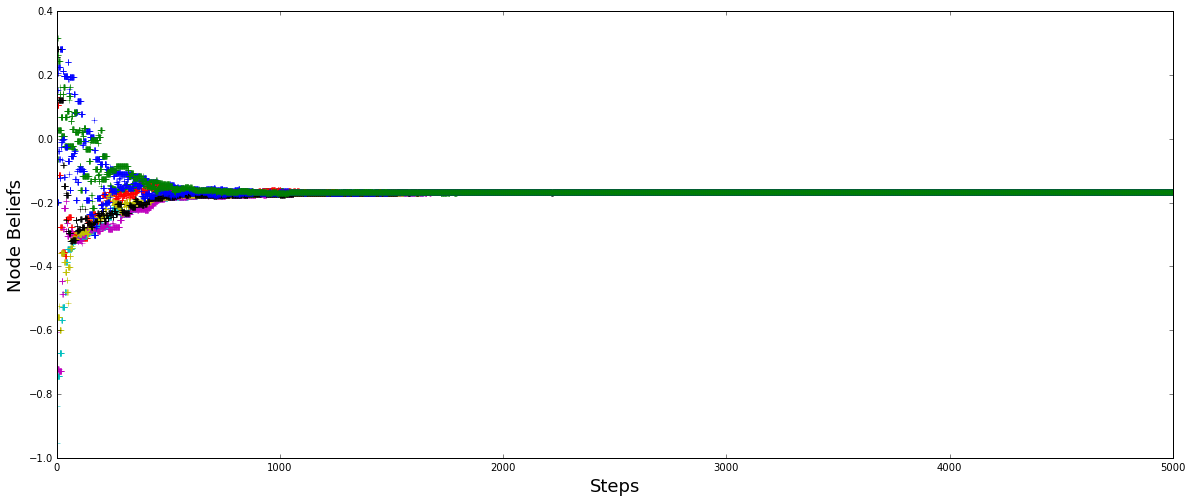}
\caption{The  social network beliefs tend to a consensus with $\beta=0.2$. }
\label{fig:unbalance0.2}
\end{center}
\end{figure}

\begin{figure}[t]
\begin{center}
\includegraphics[width=4.8in]{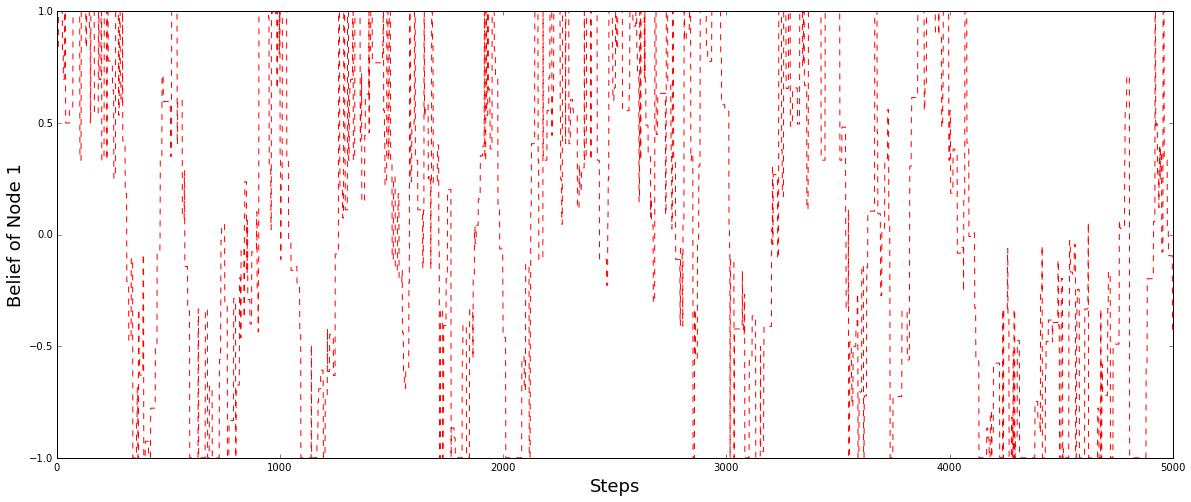}
\caption{The belief oscillation for a particular node  with $\beta=7$. }
\label{fig:unbalance7}
\end{center}
\end{figure}

Next, we select a social graph which is   neither strongly nor weakly balanced, as in Figure \ref{fig:notbalancegraph}. In Figure \ref{fig:unbalance0.2}, we plot one of the typical sample paths of the random belief evolution with $\beta=0.2$, where clearly belief consensus is achieved.    In Figure \ref{fig:unbalance7}, we plot one of the typical sample paths of the random belief evolution for a selected node with $\beta=7$, where the node belief alternatively  touches the two boundaries $-1$ and $1$ in the plotted $5000$ steps.

These numerical results are consistent with the results in Theorems \ref{strongbalance}, \ref{thmweakbalance}, and \ref{thmergodic}.

\section{Conclusions}

The  evolution of opinions over signed  social networks was studied. Each link marking interpersonal interaction in the network was associated with a sign indicating friend or enemy relations. The dynamics of opinions were defined along  positive and negative links, respectively. We have presented a comprehensive analysis to the belief convergence and divergence under various modes: in expectation, in mean-square, and almost surely. Phase transitions were established with sharp thresholds for the mean and mean-square convergence. In the almost sure sense, some surprising results were presented. When opinions have hard lower and upper bounds with asymmetric updates, the classical structure balance properties  were shown to play a key role  in the belief clustering.
We believe that these results have largely extended our understanding to how trustful and antagonistic relations shape  social opinions.

 Some interesting directions for future research include the following topics. Intuitively there is  some natural   coupling between  the structure dynamics  and the opinion evolution for  signed networks. How this coupling determines the formation of the social structure is an interesting question bridging the studies on the dynamics of signed graphs (e.g., \cite{continuouspnas}) and the opinion dynamics on signed social networks (e.g., \cite{altafini1,altafini2}). It will also be interesting to ask what might be a proper model, and what  the role of structure balance is,   for Bayesian opinion evolution on signed social networks (e.g., \cite{baysian2}).

\vspace{5mm}

\section*{Appendix: Proofs of Statements}

\subsection*{A. The Triangle Lemma}

We establish a key technical lemma on the relative beliefs of three nodes in the network in the presence of at least one link among the three nodes. Denote ${J}_{ab}(k):= |x_{a}(k)-x_{b}(k)|$ for $a,b\in \mathsf{V}$ and $k\geq0$.

\begin{lemma}\label{lemmatriangle}  Let $i_0, i_1, i_2$ be three different nodes in $\mathsf{V}$. Suppose $\{i_0,i_1\}\in \mathsf{E}$. There  exist  a positive number $\delta>0$ and an integer $Z>0$,  such that
\begin{itemize}
\item[(i)]   there is  a sequence of $Z$ successive  node pairs  leading to ${J}_{i_1i_2}(Z)\geq\delta {J}_{i_0i_1}(0)$;
    \item[(ii)] there is  a sequence of $Z$ successive  node pairs leading to ${J}_{i_1i_2}(Z)\geq\delta {J}_{i_0i_2}(0)$.
\end{itemize}
Here $\delta$ and $Z$ are absolute constants in the sense that they do not depend on $i_0,i_1,i_2$, nor on the values held at these nodes.
\end{lemma}
{\it Proof.} We assume $n\geq5$. Generality is not lost  by making this assumption because for  $n=3$ and $n=4$, some (tedious but straightforward) analysis on each possible  $\mathsf{G}$ leads to the desired conclusion.

{\it (i)}. There are two cases: $\{i_0,i_1\}\in \mathsf{E}_{\rm pst}$, or  $\{i_0,i_1\}\in \mathsf{E}_{\rm neg}$. We prove the desired conclusion  for each of the two cases.  Without loss of generality, we  assume that $x_{i_0}(0)<x_{i_1}(0)$.
\begin{quotation}

\begin{itemize}
\item Let $\{i_0,i_1\}\in \mathsf{E}_{\rm pst}$. If $x_{i_2}(0)\in \big[\frac{3}{4}x_{i_0}(0)+\frac{1}{4}x_{i_1}(0),\frac{1}{4}x_{i_0}(0)+\frac{3}{4}x_{i_1}(0)\big]$, we have $J_{i_1i_2}(0)\geq \frac{1}{4}J_{i_0i_1}(0) $. Thus, the desired conclusion holds for $\delta=\frac{1}{4}$, arbitrary $Z>0$, and any node pair sequence over $0,1,\dots,Z-1$ for which $i_0,i_1,i_2$ are never selected.

    On the other hand suppose $x_{i_2}(0)\notin \big[\frac{3}{4}x_{i_0}(0)+\frac{1}{4}x_{i_1}(0),\frac{1}{4}x_{i_0}(0)+\frac{3}{4}x_{i_1}(0)\big]$.  Take
    \begin{align}\label{10}
 d_\ast=\begin{cases} \lceil\log_{|1-2\alpha|} \frac{1}{4}\rceil& {\rm if}\ \alpha\neq \frac{1}{2},\\
   1, & {\rm if}\ \alpha=\frac{1}{2}.
   \end{cases}
   \end{align}
   If $\{i_0,i_1\}$ is selected for $0,1,\dots,d_\ast-1$, we obtain  $J_{i_0i_1}(d_\ast)\leq \frac{1}{4}  J_{i_0i_1}(0)$
which leads to
$$
x_{i_1}(d_\ast)\in \Big[\frac{5}{8}x_{i_0}(0)+\frac{2}{8}x_{i_1}(0),\frac{3}{8}x_{i_0}(0)+\frac{5}{8}x_{i_1}(0)\Big]; \ \ x_{i_2}(d_\ast)=x_{i_2}(0).
$$
This gives us $J_{i_1i_2}(d_\ast)\geq \frac{1}{8}J_{i_0i_1}(0)$.

\item Let $\{i_0,i_1\}\in \mathsf{E}_{\rm neg}$.  If $x_{i_2}(0)\notin \big[\frac{1}{2}x_{i_0}(0)+\frac{1}{2}x_{i_1}(0), -\frac{1}{2}x_{i_0}(0)+\frac{3}{2}x_{i_1}(0)\big]$, we  have $J_{i_1i_2}(0)\geq \frac{1}{2}J_{i_0i_1}(0) $. The conclusion holds for $\delta=\frac{1}{2}$, arbitrary $Z>0$, and any node pair sequence over $0,1,\dots,Z-1$ for which $i_0,i_1,i_2$ are never selected.

    On the other hand let $x_{i_2}(0)\in \big[\frac{1}{2}x_{i_0}(0)+\frac{1}{2}x_{i_1}(0), -\frac{1}{2}x_{i_0}(0)+\frac{3}{2}x_{i_1}(0)\big]$. Take ${d}^\ast= \lceil \log_{1+2\beta} 4\rceil$. Let  $\{i_0,i_1\}$ be selected for $0,1,\dots,{d}^\ast-1$. In this case,  $x_{i_0}(s)$ and $x_{i_1}(s)$ are  symmetric with respect to their center $\frac{1}{2}x_{i_0}(0)+\frac{1}{2} x_{i_1}(0)$ for all $s=0,\dots,d^\ast$, and
    $J_{i_0i_1}(d_\ast) \geq 4J_{i_0i_1}(0)$. Thus we have $ x_{i_2}({d}^\ast)= x_{i_2}(0)$, and
\begin{align}
x_{i_1}({d}^\ast)&\geq \frac{1}{2}x_{i_0}(0) +\frac{1}{2}x_{i_1}(0) +2(x_{i_1}(0)-x_{i_0}(0))\nonumber\\
&=-\frac{3}{2}x_{i_0}(0) +\frac{5}{2}x_{i_1}(0).
\end{align}
We can therefore conclude that $J_{i_1i_2}({d}^\ast) \geq J_{i_0i_1}(0)$.
\end{itemize}
\end{quotation}
In summary, the desired conclusion holds for $\delta=\frac{1}{8}$ and \begin{align}\label{11}
 Z=\begin{cases} \max\{ \lceil \log_{1+2\beta} 4\rceil,  \lceil \log_{|1-2\alpha|} \frac{1}{4}\rceil\} & {\rm if}\ \alpha\neq \frac{1}{2}\\
   \lceil \log_{1+2\beta} 4\rceil, & {\rm if}\ \alpha=\frac{1}{2}.
   \end{cases}
\end{align}

\noindent {\it (ii)}. We distinguish  the cases $\{i_0,i_1\}\in \mathsf{E}_{\rm pst}$ and  $\{i_0,i_1\}\in \mathsf{E}_{\rm neg}$. Without loss of generality, we  assume that $x_{i_0}(0)<x_{i_2}(0)$.
\begin{quotation}

\begin{itemize}
\item Let $\{i_0,i_1\}\in \mathsf{E}_{\rm pst}$. If  $x_{i_1}(0)\notin \big[\frac{1}{2}x_{i_0}(0)+\frac{1}{2}x_{i_2}(0), -\frac{1}{2}x_{i_0}(0)+\frac{3}{2}x_{i_2}(0)\big]$, we have $J_{i_1i_2}(0)\geq \frac{1}{2}J_{i_0i_2}(0) $. The conclusion holds for $\delta=\frac{1}{2}$, arbitrary $Z>0$, and any node pair sequence $0,1,\dots,Z-1$ for which $i_0,i_1,i_2$ are never selected.

    Now let  $x_{i_1}(0)\in \big[\frac{1}{2}x_{i_0}(0)+\frac{1}{2}x_{i_2}(0), -\frac{1}{2}x_{i_0}(0)+\frac{3}{2}x_{i_2}(0)\big]$. We write $x_{i_1}(0)=(1-\varsigma)x_{i_0}(0)+\varsigma x_{i_2}(0)$ with $\varsigma\in[\frac{1}{2},\frac{3}{2}]$. Let $\{i_0,i_1\}$ be the node pair selected for $0,1,\dots,d_\ast-1$ with $d_\ast$  defined by (\ref{10}). Note that according to the structure of the update rule,  $x_{i_0}(s)$ and $x_{i_1}(s)$ will be symmetric with respect to their center $(1-\frac{\varsigma}{2})x_{i_0}(0)+\frac{\varsigma}{2} x_{i_2}(0)$ for all $s=0,\dots,d_\ast$, and
    $J_{i_0i_1}(d_\ast) \leq \frac{1}{4} J_{i_0i_1}(0)$. This gives us  $x_{i_2}(d_\ast)=x_{i_2}(0)$ and
  \begin{align}
    x_{i_1}(d_\ast)&\in \Big[(1-\frac{\varsigma}{2})x_{i_0}(0)+\frac{\varsigma}{2} x_{i_2}(0)-\frac{1}{8}(x_{i_1}(0)-x_{i_0}(0)), \nonumber\\
    &\ \ (1-\frac{\varsigma}{2})x_{i_0}(0)+\frac{\varsigma}{2} x_{i_2}(0)+\frac{1}{8}(x_{i_1}(0)-x_{i_0}(0))\Big]\nonumber\\
    &=\Big[(1-\frac{3\varsigma}{8})x_{i_0}(0)+\frac{3\varsigma}{8} x_{i_2}(0),(1-\frac{5\varsigma}{8})x_{i_0}(0)+\frac{5\varsigma}{8} x_{i_2}(0)\Big],
\end{align}
which implies
\begin{align}
J_{i_1i_2}(d_\ast) \geq  (1-\frac{5\varsigma}{8})J_{i_0i_2}(0)\geq \frac{1}{16}J_{i_0i_2}(0).
\end{align}

\item Let $\{i_0,i_1\}\in \mathsf{E}_{\rm neg}$.  If $x_{i_1}(0)\notin \big[\frac{1}{2}x_{i_0}(0)+\frac{1}{2}x_{i_2}(0), -\frac{1}{2}x_{i_0}(0)+\frac{3}{2}x_{i_2}(0)\big]$, the conclusion holds for the same reason as in the case where  $\{i_0,i_1\}\in \mathsf{E}_{\rm pst}$.

      Now let  $x_{i_1}(0)\in \big[\frac{1}{2}x_{i_0}(0)+\frac{1}{2}x_{i_2}(0), -\frac{1}{2}x_{i_0}(0)+\frac{3}{2}x_{i_2}(0)\big]$. We continue to use the notation  $x_{i_1}(0)=(1-\varsigma)x_{i_0}(0)+\varsigma x_{i_2}(0)$ with $\varsigma\in[\frac{1}{2},\frac{3}{2}]$. Let $\{i_0,i_1\}$ be the node pair selected for $0,1,\dots,d^\ast-1$  where ${d}^\ast= \lceil \log_{1+2\beta} 4\rceil$. In this case,  $x_{i_0}(s)$ and $x_{i_1}(s)$ are still symmetric with respect to their center $(1-\frac{\varsigma}{2})x_{i_0}(0)+\frac{\varsigma}{2} x_{i_2}(0)$ for all $s=0,1,\dots,d^\ast$, and
    $J_{i_0i_1}(d_\ast) \geq 4J_{i_0i_1}(0)$. This gives us  $x_{i_2}(d_\ast)=x_{i_2}(0)$ and
  \begin{align}
    x_{i_1}(d_\ast)&\geq (1-\frac{\varsigma}{2})x_{i_0}(0)+\frac{\varsigma}{2} x_{i_2}(0)+{2}(x_{i_1}(0)-x_{i_0}(0)) \nonumber\\
    &=(1-\frac{5\varsigma}{2})x_{i_0}(0)+\frac{5\varsigma}{2} x_{i_2}(0)
\end{align}
which implies
\begin{align}
J_{i_1i_2}(d_\ast) \geq  (\frac{5\varsigma}{2}-1)J_{i_0i_2}(0)\geq \frac{1}{4}J_{i_0i_2}(0).
\end{align}
\end{itemize}
\end{quotation}
In summary, the desired conclusion holds for $\delta=\frac{1}{16}$ with $Z$ defined in (\ref{11}).

\subsection*{B. Proof of Theorem \ref{nosurvivor}}

Introduce
$$
{X}_{\rm min}(k)=\min_{i\in \mathsf{V}} x_i(k); \quad{X}_{\rm max}(k)=\max_{i\in \mathsf{V}} x_i(k).
$$
We define $\mathfrak{X}(k)={X}_{\rm max}(k)-{X}_{\rm min}(k)$. Suppose belief divergence is achieved almost surely. Take a constant $N_0$ such that $N_0>\mathfrak{X}(0)$. Then almost surely,
$$
{K}_1:=\inf_k \{\mathfrak{X}(k)\geq N_0\}
$$
is a finite number. Then ${K}_1$ is a {\it stopping time} for the node pair selection process ${G}_k, k=0,1,2,\dots$ since
$$
\{{K}_1 =k\} \in \sigma({G}_0,\dots,{G}_{k-1})
$$
for all $k=1,2,\dots$ due to the fact that  $\mathfrak{X}(k)$ is, indeed, a function of ${G}_0,\dots,{G}_{k-1}$. Strong Markov Property leads to:
 ${G}_{{K}_1},{G}_{{K}_1+1},\dots$ are independent of $\mathcal{F}_{{K}_1-1}$, and they are i.i.d. with the same distribution as ${G}_0$ (e.g., Theorem 4.1.3 in \cite{durr}).

Now take two different (deterministic) nodes $i_0$ and $j_0$. Since $\mathfrak{X}({K}_1)\geq N_0$, there must be two different (random) nodes $i_\ast$ and $j_\ast$ satisfying $x_{i_\ast}({K}_1)<x_{j_\ast}({K}_1)$ with $J_{i_\ast j_\ast}({K}_1)\geq N_0$. We make the following claim.

{\it Claim.} There  exist  a positive number $\delta_0>0$ and an integer $Z_0>0$ ($\delta_0$ and $Z_0$ are deterministic constants) such that we  can always select  a sequence of node pairs  for  time steps ${K}_1,{K}_1+1,{K}_1+Z_0-1$ which guarantees ${J}_{i_0j_0}({K}_1+Z_0)\geq\delta_0 N_0$.

First of all note that $i_\ast$ and $j_\ast$ are independent with $G_{{K}_1},G_{{K}_1+1},\dots,$ since $i_\ast, j_\ast\in \mathcal{F}_{{K}_1-1}$. Therefore, we can treat $i_\ast$ and $j_\ast$ as deterministic  and prove the claim for all choices of such $i_\ast$ and $j_\ast$ (because we can always carry out the analysis conditioned on different events $\{i_\ast=i,j_\ast=j\}$, $i,j\in\mathsf{V}$). We proceed  the proof recursively taking advantage of the Triangle Lemma.

 Suppose $\{i_0,j_0\}=\{i_\ast,j_\ast\}$, the claim holds trivially.
 Now suppose $i_0\notin \{i_\ast,j_\ast\}$. Either $J_{i_0 i_\ast}({K}_1)\geq \frac{N_0}{2}$ or $J_{i_0 j_\ast}({K}_1)\geq \frac{N_0}{2}$ must hold. Without loss of generality we assume $J_{i_0 i_\ast}({K}_1)\geq \frac{N_0}{2}$. Since $\mathsf{G}$ is connected, there is a path $i_0 i_1\dots i_{\tau}j_0$ in $\mathsf{G}$ with $\tau\leq n-2$.

     Based on Lemma \ref{lemmatriangle}, there exist $\delta>0$ and integer $Z>0$ such that a selection of node pair sequence for ${K}_1,{K}_1+1,\dots,{K}_1+Z-1$ leads to
     $$
     J_{i_0i_1}({K}_1+Z)\geq \delta J_{i_0 i_\ast}({K}_1)\geq \frac{\delta N_0}{2}
     $$
     since $\{i_0, i_1\}\in \mathsf{E}$. Applying recursively the Triangle Lemma based on the fact that $\{i_1,i_2\},\dots, \{i_\tau,j_0\} \in \mathsf{E}$,  we see that  a selection of node pair sequence for ${K}_1,{K}_1+1,\dots,{K}_1+(\tau+1)Z-1$ will give us
     $$
     J_{i_0j_0}({K}_1+(\tau+1)Z)\geq \delta^{\tau+1} J_{i_0 i_\ast}({K}_1)\geq \frac{\delta^{\tau+1} N_0}{2}.
     $$
Since $\tau\leq n-2$, the claim always holds for $\delta_0=\frac{\delta^{n-1} }{2}$ and $Z_0 = (n-1)Z$, independently  of $i_\ast$ and $j_\ast$.

Therefore, denoting $p_\ast=\min\{ p_{ij}+p_{ji}:\{i,j\}\in\mathsf{E}\}$, the claim we just proved yields that
\begin{align}
\mathbb{P}\Big(  J_{i_0j_0}({K}_1+(n-1)Z)\geq  \frac{\delta^{n-1} N_0}{2}\Big)\geq \Big(\frac{p_\ast}{n}\Big)^{(n-1)Z}.
\end{align}

We proceed the analysis by recursively defining
$$
{K}_{m+1}:=\inf \big\{k\geq {K}_{m}+Z_0:\mathfrak{X}(k)\geq N_0\big\},\ \ m=1,2,\dots.
$$
Given that belief divergence is achieved, ${K}_{m}$ is finite for all $m\geq 1$ almost surely. Thus,
\begin{align}
\mathbb{P}\Big(  J_{i_0j_0}({K}_m+Z_0)\geq  \frac{\delta^{n-1} N_0}{2}\Big)\geq \Big(\frac{p_\ast}{n}\Big)^{Z_0},
\end{align}
for all $m=1,2,\dots$.
Moreover, the node pair sequence
$$
{G}_{{K}_{1}},\dots,{G}_{{K}_{1}+Z_0-1};\dots \dots;{G}_{{K}_{m}},\dots,{G}_{{K}_{m}+Z_0-1};\dots \dots
$$
are independent and have the same distribution as ${G}_0$ (This is due to that $\mathcal{F}_{{K}_{1}}\subseteq \mathcal{F}_{{K}_{1}+1} \subseteq \dots \subseteq \mathcal{F}_{{K}_{1}+Z_0-1}\subseteq \mathcal{F}_{{K}_{2}}\dots$. (cf. Theorem 4.1.4 in \cite{durr})).

Therefore, we can finally  invoke  the second Borel-Cantelli Lemma (cf. Theorem 2.3.6 in \cite{durr}) to conclude that almost surely, there exists an infinite subsequence ${K}_{m_s},s=1,2,\dots,$ satisfying
\begin{align}\label{r12}
 J_{i_0j_0}({K}_{m_s}+Z_0)\geq  \frac{\delta^{n-1} N_0}{2},\ \ s=1,2,\dots,
\end{align}
conditioned on that belief divergence is achieved. Since $\delta$ is a constant and $N_0$ is arbitrarily chosen, (\ref{r12}) is equivalent to $\mathbb{P}\big(\limsup_{k\rightarrow \infty}  \big |x_{i_0}(k)-x_{j_0}(k)\big|=\infty\big)=1$, which completes  the proof.

\subsection*{C. Proof of Lemma \ref{thmliveordie}}

{(i).} It suffices to show that $\mathbb{P}\big(\limsup_{k\rightarrow \infty} \mathfrak{X}(k)\in [a_\ast,b_\ast]  \big)=0$ for all $0<a_\ast<b_\ast$. We prove the statement by  contradiction. Suppose $\mathbb{P}\big(\limsup_{k\rightarrow \infty} \mathfrak{X}(k)\in [a_\ast,b_\ast]  \big)=p>0$ for some $0<a_\ast<b_\ast$.

Take $0<\varepsilon<1$ and define $a=a_\ast(1-\varepsilon), b=b_\ast(1+\varepsilon)$. We introduce
$$
{T}_1:=\inf_k \{\mathfrak{X}(k)\in  [a, b]\}.
$$
Then ${T}_1$ is finite with probability at least $p$. ${T}_1$ is a stopping time. ${G}_{{T}_1},{G}_{{T}_1+1},\dots$ are independent of $\mathcal{F}_{{T}_1-1}$, and they are i.i.d. with the same distribution as ${G}_0$.

Now since $\mathsf{G}_{\rm neg}$ is nonempty, we take a link $\{i_\star, j_\star\}\in \mathsf{E}_{\rm neg}$. Repeating the same analysis as the proof of Theorem \ref{nosurvivor}, the following statement holds true conditioned on that ${T}_1$ is finite: { there  exist  a positive number $\delta_0>0$ and an integer $Z_0>0$ ($\delta_0$ and $Z_0$ are deterministic constants) such that we  can always select  a sequence of node pairs  for  time steps ${T}_1,{T}_1+1,{T}_1+Z_0-1$ which guarantees ${J}_{i_\star j_\star}({T}_1+Z_0)\geq\delta_0 a$.}

Here $\delta_0$ and $Z_0$ follow from the same definition in the proof of  Theorem \ref{nosurvivor}.  Take
$$
m_0= \Big\lceil \log_{2\beta +1} \frac{2b}{\delta_0 a}\Big\rceil
$$
and let $\{i_\star,j_\star\}$ be selected for ${T}_1+Z_0,\dots,{T}_1+Z_0+m_0-1$. Then noting that $\{i_\star, j_\star\}\in \mathsf{E}_{\rm neg}$, the choice of $m_0$ and the fact that ${J}_{i_\star j_\star}(s+1)=(2\beta+1){J}_{i_\star j_\star}(s)$, $s={T}_1+Z_0,\dots,{T}_1+Z_0+m_0-1$ lead to
$$
\mathfrak{X}({T}_1+Z_0+m_0) \geq {J}_{i_\star j_\star}({T}_1+Z_0+m_0) \geq(2\beta+1)^{m_0}\delta_0 a \geq 2b \geq 2b_\ast.
$$
We have proved that
\begin{align}
\mathbb{P}\Big( \mathfrak{X}({T}_1+Z_0+m_0)\geq  2b_\ast \Big| T_1<\infty\Big)\geq \Big(\frac{p_\ast}{n}\Big)^{Z_0+m_0}.
\end{align}

Similarly, we proceed the analysis by recursively defining
$$
{T}_{m+1}:=\inf \big\{k\geq {T}_{m}+Z_0+m_0:\mathfrak{X}(k)\in [a,b]\big\},\ \ m=1,2,\dots.
$$
Given $\mathbb{P}\big(\limsup_{k\rightarrow \infty} \mathfrak{X}(k)\in [a_\ast,b_\ast]  \big)=p$, ${T}_{m}$ is finite for all $m\geq 1$ with probability at least $p$. Thus, there holds
\begin{align}
\mathbb{P}\Big( \mathfrak{X}({T}_m+Z_0+m_0)\geq  2b_\ast\Big| T_m<\infty\Big)\geq \Big(\frac{p_\ast}{n}\Big)^{Z_0+m_0},\ \ m=1,2,\dots.
\end{align}
 The independence of
$$
{G}_{{T}_{1}},\dots,{G}_{{T}_{1}+Z_0+m_0-1};\dots \dots;{G}_{{T}_{m}},\dots,{G}_{{T}_{m}+Z_0+m_0-1};\dots \dots$$
once again allows us to invoke  the Borel-Cantelli Lemma to conclude that almost surely, there exists an infinite subsequence ${T}_{m_s},s=1,2,\dots,$ satisfying
\begin{align}
\mathfrak{X}( {T}_{m_s}+Z_0+m_0) \geq 2b_\ast  ,\ \ s=1,2,\dots,
\end{align}
given  that ${T}_m, m=1,2\dots$, are finite. In other words, we have obtained that
\begin{align}\label{12}
\mathbb{P}\Big(\limsup_{k\rightarrow \infty}\mathfrak{X}(k) \geq 2b_\ast \Big|\limsup_{k\rightarrow \infty}\mathfrak{X}(k) \in[a_\ast,b_\ast]\Big)=1,
\end{align}
which is impossible and the first part of the theorem has been proved.

\vspace{2mm}

\noindent{(ii).} It suffices to show that $\mathbb{P}\big(\liminf_{k\rightarrow \infty} \mathfrak{X}(k)\in [a_\ast,b_\ast]  \big)=0$ for all $0<a_\ast<b_\ast$. The proof is again by contradiction. Assume that $\mathbb{P}\big(\liminf_{k\rightarrow \infty} \mathfrak{X}(k)\in [a_\ast,b_\ast]  \big)=q>0$. Let $a, b$, and ${T}_1:=\inf_k \{\mathfrak{X}(k)\in  [a,b]\}$ as defined earlier. ${T}_1$ is finite with probability at least $q$.

Let $\ell_0 \in \mathsf{V}$ satisfying $x_{\ell_0}({T}_1)={X}_{\rm min} ({T}_1)$. There is a path from $\ell_0$ to every other node in the network since $\mathsf{G}_{\rm pst }$ is connected. We introduce
$$
{\mathsf{V}}^\dag_t:=\{j:d(\ell_0,j)=t\ {\rm in}\ \mathsf{G}_{\rm pst }\},\ t=0,\dots,{\rm diam}(\mathsf{G}_{\rm pst })
$$
as a partition of $\mathsf{V}$.
We relabel the nodes in $\mathsf{V}\setminus \{\ell_0\}$ in the following manner.
\begin{align}
& \ell_s \in {\mathsf{V}}^\dag_1,s=1,\dots,|{\mathsf{V}}^\dag_1|;\nonumber\\
& \ell_s \in {\mathsf{V}}^\dag_2, s=|{\mathsf{V}}^\dag_1|+1,\dots,|{\mathsf{V}}^\dag_1|+|{\mathsf{V}}^\dag_2|;\nonumber\\
 &\dots \dots \nonumber\\
 & \ell_s\in{\mathsf{V}}^\dag_{{\rm diam}(\mathsf{G}_{\rm pst })},s=\sum_{t=1}^{{\rm diam}(\mathsf{G}_{\rm pst })-1}|{\mathsf{V}}^\dag_t|,\dots, n-1. \nonumber
\end{align}
Then the definition of ${\mathsf{V}}^\dag_t$ and the connectivity  of $\mathsf{G}_{\rm pst }$ allow us to select a sequence of node pairs in the form of
$$
{G}_{{T}_1+s}=\{\ell_\rho, \ell_{s+1}\}, \ \  \{\ell_\rho, \ell_{s+1}\}\in \mathsf{E}_{\rm pst }\ {\rm with}\ \rho\leq s,
$$
for $s=0,\dots,n-2$. Next we give an estimation for $\mathfrak{X}$ under the selected sequence of node pairs.
\begin{quotation}
\begin{itemize}
\item Since $\{\ell_0,\ell_1\}$ is selected at time ${T}_1$, we have
\begin{align}
x_{\ell_0}({T}_1+1)&=(1-\alpha)x_{\ell_0}({T}_1)+\alpha x_{\ell_1}({T}_1)\leq(1-\alpha){X}_{\rm min}({T}_1)+\alpha{X}_{\rm max}({T}_1);\nonumber\\
x_{\ell_1}({T}_1+1)&=(1-\alpha)x_{\ell_1}({T}_1)+\alpha x_{\ell_0}({T}_1)\leq(1-\alpha){X}_{\rm max}({T}_1)+\alpha {X}_{\rm min}({T}_1).
\end{align}
This leads to  $x_{\ell_s}({T}_1+1)\leq(1-\alpha_\ast){X}_{\rm min}({T}_1)+\alpha_\ast{X}_{\rm max}({T}_1),s=0,1$, where $\alpha_\ast=\max\{\alpha,1-\alpha\}$.

\item Note that ${X}_{\rm max}({T}_1+1)={X}_{\rm max}({T}_1)$, and that either  $\{\ell_0,\ell_2\}$ or  $\{\ell_1,\ell_2\}$ is selected at time ${T}_1+1$. We deduce:
\begin{align}
x_{\ell_s}({T}_1+2)& \leq(1-\alpha)\big[(1-\alpha_\ast){X}_{\rm min}({T}_1)+\alpha_\ast{X}_{\rm max}({T}_1)\big]+\alpha{X}_{\rm max}({T}_1)\nonumber\\
&\leq  (1-\alpha_\ast)^2{X}_{\rm min}({T}_1)+\big(1-(1-\alpha_\ast)^2\big){X}_{\rm max}({T}_1), \ \ s=0, 1; \nonumber\\
x_{\ell_2}({T}_1+2)&\leq \alpha[(1-\alpha_\ast){X}_{\rm min}({T}_1)+\alpha_\ast{X}_{\rm max}({T}_1)]+(1-\alpha){X}_{\rm max}({T}_1)\nonumber\\
&\leq  (1-\alpha_\ast)^2{X}_{\rm min}({T}_1)+\big(1-(1-\alpha_\ast)^2\big){X}_{\rm max}({T}_1),
\end{align}
Thus we obtain  $x_{\ell_s}({T}_1+2)\leq(1-\alpha_\ast)^2{X}_{\rm min}({T}_1)+\big(1-(1-\alpha_\ast)^2\big){X}_{\rm max}({T}_1),s=0,1,2$.
\item We carry on the analysis recursively, and finally get:
 $$
 x_{\ell_s}({T}_1+n-1)\leq(1-\alpha_\ast)^{n-1}{X}_{\rm min}({T}_1)+\big(1-(1-\alpha_\ast)^{n-1}\big){X}_{\rm max}({T}_1),
 $$
 for $s=0,1,2,\dots,n-1$. Equivalently:
\begin{align}\label{119}
{X}_{\rm max}({T}_1+n-1)\leq(1-\alpha_\ast)^{n-1}{X}_{\rm min}({T}_1)+\big(1-(1-\alpha_\ast)^{n-1}\big){X}_{\rm max}({T}_1).
\end{align}
We conclude that:
 \begin{align}
\mathfrak{X}({T}_1+n-1)&={X}_{\rm max}({T}_1+n-1)-{X}_{\rm min}({T}_1+n-1)\nonumber\\
&={X}_{\rm max}({T}_1+n-1)-{X}_{\rm min}({T}_1)\nonumber\\
&\leq  r_0 \mathfrak{X}({T}_1),
\end{align}
where $r_0=1-(1-\alpha_\ast)^{n-1}$ is a constant in $(0,1)$.
\end{itemize}
\end{quotation}

With the above analysis taking
$$
L_0= \Big \lceil \log_{r_0} \frac{a}{2b}\Big\rceil,
$$
and selecting the given pair sequence periodically for  $L_0$ rounds, we obtain
\begin{align}\label{16}
\mathfrak{X}({T}_1+(n-1)L_0)\leq  r_0^{L_0} \mathfrak{X}({T}_1) \leq\frac{a}{2b} \cdot b =\frac{a}{2}< \frac{a_\ast}{2}.
\end{align}
In light of (\ref{16}) and the selection of the node pair sequence, we have obtained that
\begin{align}
\mathbb{P}\Big( \mathfrak{X}({T}_1+(n-1)L_0)\leq \frac{a_\ast}{2}\Big)\geq \Big(\frac{p_\ast}{n}\Big)^{(n-1)L_0}
\end{align}
given  that ${T}_1$ is finite. We repeat the above argument for ${T}_{m+1}$, $m=2,3\dots$. Borel-Cantelli Lemma then implies
\begin{align}
\mathbb{P}\Big(\liminf_{k\rightarrow \infty}\mathfrak{X}(k) \leq \frac{a_\ast}{2}\ \Big|\liminf_{k\rightarrow \infty}\mathfrak{X}(k) \in[a_\ast,b_\ast]\Big)=1,
\end{align}
which is impossible and  completes the proof.

\subsection*{D. Proof of Theorem \ref{thmzeroone}}

Let $\omega\notin {\mathscr{C}} $. Then there exists an initial value $x^0\in \mathds{R}^n$ from which
\begin{align}\label{17}
\limsup_{k\rightarrow \infty} \mathfrak{X}(k)(\omega)>0.
\end{align}
According to Lemma \ref{thmliveordie}, (\ref{17}) implies that
\begin{align}
\mathbb{P} \Big( \limsup_{k\rightarrow \infty} \mathfrak{X}(k)=\infty \Big|\mathscr{C}^c\Big)=\mathbb{P} \Big( \mathscr{D} \big|\mathscr{C}^c\Big)=1,
\end{align}
which implies  $\mathbb{P}(\mathscr{C})+\mathbb{P}(\mathscr{D})=1$.

With $\mathbb{P}(\mathscr{C})+\mathbb{P}(\mathscr{D})=1$, $\mathscr{D}$ is a trivial event as long as   $\mathscr{C}$ is a trivial event. Therefore, for completing the proof we just need to verify that $\mathscr{C}$ is a trivial event.

We first show that $
\mathscr{C}= \big\{\lim_{k\rightarrow \infty} W_k\dots W_0 =U  \big\}$.  In fact, if  $\limsup_{k\rightarrow \infty} \max_{i,j}|x_i(k)-x_j(k)|=0$  under  $x^0\in\mathds{R}^n$, then we have $\lim_{k\rightarrow \infty}x(k)=\frac{1}{n}\mathbf{1}\mathbf{1}' x^0$ because the sum of the beliefs is preserved. Therefore, we can restrict the analysis to $x^0=e_i$, $i=1,\dots,n$ and on can readily see that $\mathscr{C}= \big\{\lim_{k\rightarrow \infty} W_k\dots W_0 =U  \big\}$.

Next, we apply the  argument,  which was originally introduced in \cite{jad08} for establishing  the weak ergodicity of product of random stochastic matrices with positive diagonal terms, to conclude that $\mathscr{C}$ is a trivial event. A more general treatment to zero-one laws of random averaging algorithms can be found in \cite{touri}. Define a sequence of event $\mathscr{C}_s= \big\{\lim_{k\rightarrow \infty} W_k\dots W_s =U  \big\}$ for $s=1,2,\dots$. We see that \begin{itemize}
\item $\mathbb{P}(\mathscr{C}_s)=\mathbb{P}(\mathscr{C})$ for all $s=1,2,\dots$ since $W_k,k=0,1,\dots,$ are i.i.d.
\item  $\mathscr{C}_{s+1}\subseteq \mathscr{C}_{s}$ for all $s=1,2,\dots$ since $\lim_{k\rightarrow \infty} W_k\dots W_{s+1}=U$ implies $\lim_{k\rightarrow \infty} W_k\dots W_s =U$ due to the fact that $UW_s \equiv U$.
\end{itemize}
Therefore, we have $\bigcap_{s=1}^\infty \mathscr{C}_{s}$ is a tail event within the tail $\sigma$-field $\bigcap_{s=1}^\infty \sigma({G}_s,{G}_{s+1},\dots)$. By Kolmogorov's zero-one law, $\bigcap_{s=1}^\infty \mathscr{C}_{s}$ is a trivial event. Hence $\mathbb{P}(\mathscr{C})=\lim_{s\rightarrow \infty} \mathbb{P}( \mathscr{C}_{s})=\mathbb{P}(\bigcap_{s=1}^\infty \mathscr{C}_{s})$ is a trivial event, and the desired conclusion follows.

\subsection*{E. Proof of Theorem \ref{thmphasetran}}

Theorem \ref{thmphasetran} is a direct consequence of the following    lemmas.

\begin{lemma}\label{lem2}
Suppose $\mathsf{G}_{\rm pst}$ is connected. Then for every fixed  $\alpha\in(0,1)$, we have $\mathbb{P}(\mathscr{C})=1$ for all  $0\leq \beta<\beta^\natural$ with
$$
\beta^\natural:=\sup\Big\{\beta:\ \beta(1+\beta)< \frac{\lambda_2( L_{\rm pst}^\dag)} {\lambda_{\rm max}( L_{\rm neg}^\dag)}\alpha(1-\alpha)\Big\}.
$$
\end{lemma}
{\it Proof.} Let $x_{\rm ave}=\sum_{i\in \mathsf{V}} x_i(0)/n$ be the average of the initial beliefs. We introduce $V(k)=\sum_{i=1}^n |x_i(k)-x_{\rm ave}|^2= \big|(I-U)x(k)\big|^2$. The evolution of $V(k)$ follows from
\begin{align}\label{20}
\mathbb{E} \Big\{ V(k+1)  \Big| x(k)\Big\}
&=\mathbb{E}\Big\{ x(k+1)'  (I-U)^2x(k+1)  \Big |x(k) \Big\}\nonumber\\
&\stackrel{a)}{=}\mathbb{E}\Big\{ x(k)' W(k)  (I-U) W(k)x(k)  \Big | x(k) \Big\}\nonumber\\
&\stackrel{b)}{=}\mathbb{E}\Big\{ x(k)'(I-U) \big[ W(k)  (I-U) W(k)\big](I-U)x(k)  \Big |x(k) \Big\}\nonumber\\
&\stackrel{c)}{\leq}\lambda_{\rm max}\big( \mathbb{E} \{  W(k)  (I-U) W(k)\} \big) \big| (I-U)x(k)\big|^2 \nonumber\\
&\stackrel{d)}{=}\lambda_{\rm max}\big( \mathbb{E}\{ W^2(k)\}-U \big) V(k),
\end{align}
where $a)$ is based on the facts that $W(k)$ is symmetric and the simple fact $(I-U)^2=I-U$, $b)$ holds because $(I-U)W(k)=W(k)(I-U)$ always holds and again $(I-U)^2=I-U$, $c)$ follows from Rayleigh-Ritz theorem (cf. Theorem 4.2.2 in \cite{matrix}) and the fact that $W(k)$ is independent of $x(k)$, $d)$ is based on simple algebra and $W(k)U=UW(k)=U$.

We now compute $\mathbb{E}( W^2(k))$. Note that
 \begin{align}
 &\big(I-\alpha (e_i-e_j)(e_i-e_j)'\big)^2=I-2\alpha(1-\alpha) (e_i-e_j)(e_i-e_j)'; \nonumber\\
  &\big(I+\beta (e_i-e_j)(e_i-e_j)'\big)^2=I+2\beta(1-\beta) (e_i-e_j)(e_i-e_j)'.
  \end{align}
This observation combined with (\ref{2}) leads to
\begin{align}
&\mathbb{P}\Big(W^2(k)= I-2\alpha(1-\alpha) (e_i-e_j)(e_i-e_j)'\Big)=\frac{p_{ij}+p_{ji}}{n}, \ \ \   \{i,j\}\in \mathsf{E}_{\rm pst}; \nonumber\\
&\mathbb{P}\Big(W^2(k)= I+2\beta(1+\beta)(e_i-e_j)(e_i-e_j)'\Big)=\frac{p_{ij}+p_{ji}}{n}, \ \ \   \{i,j\}\in \mathsf{E}_{\rm neg}. \nonumber
\end{align}
As a result, we have
\begin{align}
\mathbb{E} \{ W^2(k)\}=I- {2\alpha(1-\alpha)} L_{\rm pst}^\dag+ {2\beta(1+\beta )} L_{\rm neg}^\dag.
\end{align}
Consequently, we have $0<\gamma:=\lambda_{\rm max}\big( \mathbb{E}( W^2(k))-U \big)<1$ for all $\beta$ satisfying
\begin{align}\label{18}
\beta(1+\beta)< \frac{\lambda_2( L_{\rm pst}^\dag)}{\lambda_{\rm max}( L_{\rm neg}^\dag)}\alpha(1-\alpha).
\end{align}

Since $g(\beta)=\beta(1+\beta)$ is nondecreasing, we conclude from (\ref{20}) that
\begin{align}\label{21}
\mathbb{E} \big\{ V(k+1)  \big| x(k)\big\} < \gamma V(k)
\end{align}
with $0<\gamma<1$ for all $0\leq \beta<\beta^\natural$. This means that $V(k)$ is a supermartingale as long as $0\leq \gamma\leq1$ \cite{durr}, and $V(k)$ converges to a limit almost surely by the martingale convergence theorem (Theorem 5.2.9, \cite{durr}).  Next we show that this limit is zero almost surely if $0\leq\gamma<1$. Let $\epsilon >0$ and $0\leq\gamma<1$.  We have:
\begin{align}
\mathbb{P}\Big(V(k)>\epsilon\ \mbox{infinitely often}\Big)&\stackrel{a)}{=}\mathbb{P}\Big( \sum_{k=0}^\infty\mathbb{P}\big(V(k+1)>\epsilon\big| x(k)\big)=\infty\Big) \nonumber\\
&\stackrel{b)}{\leq}\mathbb{P}\Big(\frac{1} {\epsilon}\sum_{k=0}^\infty{\mathbb{E}\big\{V(k+1)\big| x(k)\big\}}=\infty\Big) \nonumber\\
&\stackrel{c)}{\leq}\mathbb{P}\Big(\frac{\gamma} {\epsilon}\sum_{k=1}^\infty  V(k)=\infty\Big),
\end{align}
where $a)$ is straightforward application of  the Second Borel-Cantelli Lemma (Theorem 5.3.2. in \cite{durr}), $b)$ is from the Markov's inequality, and $c)$ holds directly from (\ref{21}). Observing that
\begin{align}
\sum_{k=1}^\infty\mathbb{E} \{ V(k)\} \leq \sum_{k=1}^\infty \gamma^k V(0)\leq \frac{\gamma}{1-\gamma} V(0)<\infty,
\end{align}
 we obtain  $\mathbb{P}\big(\frac{\gamma} {\epsilon}\sum_{k=1}^\infty  V(k)=\infty\big)=0$. Therefore, we have proved that $\mathbb{P}\big(V(k)>\epsilon\ \mbox{infinitely often}\big)=0$, or equivalently, $\mathbb{P}(\lim_{k\rightarrow \infty}V(k)=0)=1$.

Finally, observe that:
\begin{align}
V(k)=\sum_{i=1}^n |x_i(k)-x_{\rm ave}|^2\geq |x_{\rho_1}(k)-x_{\rm ave}|^2+|x_{\rho_2}(k)-x_{\rm ave}|^2\geq \frac{1}{2}|x_{\rho_1}(k)-x_{\rho_2}(k)|^2=\frac{1}{2}\mathfrak{X}^2(k), \nonumber
\end{align}
where $\rho_1$ and $\rho_2$ are chosen such that $x_{\rho_1}(k)={X}_{\rm min}(k), x_{\rho_2}(k)={X}_{\rm max}(k)$. Hence $\mathbb{P}(\lim_{k\rightarrow \infty}V(k)=0)=1$ implies $\mathbb{P}(\lim_{k\rightarrow \infty} \mathfrak{X}(k)=0)=1$. This completes the proof. \hfill$\square$

\vspace{2mm}

\begin{remark}\label{remarkproof}
We have shown that:
\begin{align}
\mathbb{E} \big\{ V(k+1)\big\} \leq \lambda_{\rm max}\big( \mathbb{E}\{ W^2(k)\}-U \big)\mathbb{E} \big\{ V(k)\big\}
\end{align}
from (\ref{20}). A symmetric analysis leads to:
\begin{align}
\mathbb{E} \big\{ V(k+1)\big\}\geq \lambda_{\rm min}\big( \mathbb{E}\{ W^2(k)\}-U \big)\mathbb{E} \big\{ V(k)\big\}.
\end{align}
Proposition \ref{meansquarethm} readily  follows from these inequalities.
\end{remark}

\vspace{2mm}

\begin{lemma}\label{lemdivergence}
Suppose  $\alpha\in[0,1]
$ with $\alpha\neq1/2$. There exists a constant $\beta^\sharp>0$ such that  $$
\mathbb{P}(\liminf_{k\rightarrow \infty} \max_{i,j}|x_i(k)-x_j(k)|=\infty)=1
$$ for almost all initial beliefs  when  $ \beta>\beta^\sharp$.
\end{lemma}
{\it Proof.}  Suppose $\mathfrak{X}(0)>0$. We have:
   \begin{align}\label{24}
   J_{ij}(k+1)=\begin{cases}
|2\alpha -1| J_{ij}(k), & \mbox{if}\  G_k=\{i,j\}\in \mathsf{E}_{\rm pst}\\
|2\beta +1| J_{ij}(k), & \mbox{if}\  G_k=\{i,j\}\in \mathsf{E}_{\rm neg}.
\end{cases}
\end{align}
Thus, $\mathfrak{X}(k)>0$ almost surely for all $k$ as long as $\mathfrak{X}(0)>0$. As a result, the following sequence of random variables is well defined:
 \begin{align}
\zeta_k= \frac{\mathfrak{X}(k+1)}{\mathfrak{X}(k)},\ \  k=0,1,\dots.
\end{align}
The proof is based on the analysis of $\zeta_k$. We proceed in three steps.

 \noindent {Step 1.} In this step, we establish some natural upper and lower bounds for $\zeta_k$.  First of all, from (\ref{24}), it is easy to see that:
\begin{align}\label{30}
\mathbb{P}\Big(\zeta_k=\frac{\mathfrak{X}(k+1)}{\mathfrak{X}(k)}\geq|2\alpha -1| \Big)=1
\end{align}
and $\mathbb{P}\big(\zeta_k<1\big) \leq \mathbb{P}\big(\mbox{one link in $\mathsf{E}_{\rm pst}$ is selected}\big).$

On the other hand let $\{i_0,j_0\}\in \mathsf{G}_{\rm neg}$. Suppose $i_\star$ and $j_\star$ are two nodes satisfying $J_{i_\star j_\star}={\mathfrak{X}(0)}$. Repeating the analysis in the proof of Theorem \ref{nosurvivor} by recursively applying the Triangle Lemma, we conclude that there is a sequence of node pairs for time steps $0,1,\dots,(n-1)Z-1$ which guarantees
\begin{align}\label{27}
  J_{i_0j_0}((n-1)Z)\geq  \frac{\delta^{n-1}}{2}\mathfrak{X}(0)
\end{align}
where $\delta=1/16$ and $Z=\max\{ \lceil \log_{1+2\beta} 4\rceil,  \lceil \log_{|1-2\alpha|} \frac{1}{4}\rceil\}$ are defined in the Triangle Lemma. 
For the remaining of the proof we assume that $\beta$ is sufficiently large so that $\lceil \log_{1+2\beta} 4\rceil\leq  \lceil \log_{|1-2\alpha|} \frac{1}{4}\rceil$, which means that we can select $Z=\lceil \log_{|1-2\alpha|} \frac{1}{4}\rceil$ independently of $\beta$.

Now take  an integer $H_0 \geq 1$. Continuing the previous node pair sequence,  let  $\{i_0,j_0\}$ be selected at time steps $(n-1)Z,\dots,(n-1)Z+H_0-1$. It then follows from (\ref{24}) and (\ref{27}) that
\begin{align}\label{28}
 \mathfrak{X}((n-1)Z+H_0)\geq J_{i_0j_0}((n-1)Z+H_0)\geq  \frac{(2\beta+1)^{H_0}\delta^{n-1}}{2}\mathfrak{X}(0).
\end{align}
Denote $Z_{H_0}=(n-1)Z+H_0$.  This node sequence for $0,1,\dots,Z_{H_0}$, which  leads to (\ref{28}), is denoted $\mathsf{S}_{i_0j_0}([0,Z_{H_0}))$.


 \noindent {Step 2.} We now define a random variable $Q_{Z_{H_0}}(0)$, associated with the node pair selection process in steps $0,\dots,Z_{H_0}-1$, by

\begin{align}Q_{Z_{H_0}}(0)=
\begin{cases}
|2\alpha-1|^{Z_{H_0}}, & \mbox{if at least  one link in $\mathsf{E}_{\rm pst}$ is selected in steps}\ 0,1,\dots, Z_{H_0}-1;\\
\frac{(2\beta+1)^{H_0}\delta^{n-1}}{2}, & \mbox{if node sequence $\mathsf{S}_{i_0j_0}([0,Z_{H_0}))$ is selected in steps}\ 0,1,\dots, Z_{H_0}-1;\\
1,  & \mbox{otherwise}.
\end{cases}
\end{align}
In view of (\ref{30}) and (\ref{28}), we  have:
\begin{align}\label{36}
\mathbb{P}\Big(\prod_{k=0}^{Z_{H_0}-1} \zeta_{k} =\frac{ \mathfrak{X}(Z_{H_0})}{ \mathfrak{X}(0)}\geq  Q_{Z_{H_0}}(0) \Big)=1.
\end{align}

From direct calculation based on the definition of $Q_{Z_{H_0}}(0)$, we conclude that
\begin{align}\label{110}
\mathbb{E}\Big\{\log Q_{Z_{H_0}}(0)\Big\}&\geq   \Big(\frac{p_\ast}{n}\Big)^{Z_{H_0}} \log \frac{(2\beta+1)^{H_0}\delta^{n-1}}{2} + \Big(1- \big(1-\frac{p^\ast}{n})^{E_0Z_{H_0}}\Big) \log |2\alpha-1|^{Z_{H_0}}\nonumber\\
&:=C_{H_0}
\end{align}
where $p^\ast=\max\{ p_{ij}+p_{ji}:\{i,j\}\in\mathsf{E}\}$ and $E_0=|\mathsf{E}_{\rm pst}|$ denotes the number of positive links. Since $Z$ does not depend on $\beta$, we see from (\ref{110}) that for any fixed $H_0$, there is a constant $\beta^\diamondsuit (H_0)>0$  with $\lceil \log_{1+2\beta^\diamondsuit} 4\rceil\leq  \lceil \log_{|1-2\alpha|} \frac{1}{4}\rceil$ guaranteeing that
$$
\beta> \beta^\diamondsuit (H_0)\Rightarrow  C_{H_0}>0.
$$

 \noindent {Step 3.} Recursively applying the analysis in the previous steps, node pair sequences  $\mathsf{S}_{i_0j_0}([sZ_{H_0},(s+1)Z_{H_0}))$ can be found for $s=1,2,\dots,$ and $Q_{Z_{H_0}}(s),s=1,2,\dots$ can be defined associated with the node pair selection process (following the same definition of $Q_{Z_{H_0}}(0)$). Since the node pair selection process is independent of time and node states, $Q_{Z_{H_0}}(s),s=0,1,2,\dots,$ are independent random variables (not necessarily i.i.d since $\mathsf{S}_{i_0j_0}([sZ_{H_0},(s+1)Z_{H_0}))$ may correspond to different pair sequences for different $s$.) The lower bound established in (\ref{110}) holds for all $s$, i.e.,
 \begin{align}\label{40}
  \mathbb{E}\big\{\log Q_{Z_{H_0}}(s)\big\}\geq C_{H_0},s=0,1,\dots.
  \end{align}
Moreover, we can prove as (\ref{36}) was established that:
\begin{align}\label{41}
\mathbb{P}\Big(\prod_{k=0}^{tZ_{H_0}-1} \zeta_{k} =\frac{ \mathfrak{X}(tZ_{H_0})}{ \mathfrak{X}(0)}\geq  \prod_{s=0}^{t-1} Q_{Z_{H_0}}(s),\  t=0,1,2,\dots\Big)=1.
\end{align}

It is straightforward to see that $\mathbb{V}\big\{\log Q_{Z_{H_0}}(s)\big\}, s=0, 1,\dots$ is bounded uniformly in $s$. Kolmogorov's strong law of large numbers (for a sequence of mutually independent random variables under Kolmogorov criterion, see \cite{feller}) implies that:
\begin{align}\label{111}
 \mathbb{P}\Big( \lim_{t\rightarrow \infty} \frac{1}{t}\sum_{s=0}^t\Big(\log Q_{Z_{H_0}}(s) -\mathbb{E}\big\{ \log Q_{Z_{H_0}}(s)\big\}\Big) = 0\Big)=1.
\end{align}
Using (\ref{40}), (\ref{111}) further implies that:
\begin{align}\label{42}
 \mathbb{P}\Big( \liminf_{t\rightarrow \infty} \frac{1}{t}\sum_{s=0}^t \log Q_{Z_{H_0}}(s) \geq C_{H_0} \Big)=1.
\end{align}

The final part of the proof is based on (\ref{41}). With the definition of $\zeta_k$, (\ref{41}) yields:
\begin{align}
\mathbb{P}\Big( \log \mathfrak{X}\big((t+1)Z_{H_0}\big) -\log\mathfrak{X}\big(0\big)= \sum_{k=0}^{(t+1)Z_{H_0}-1} \log \zeta_{k} \geq  \sum_{s=0}^t \log Q_{Z_{H_0}}(s), \ \ t=0,1,2,\dots\Big)=1,\nonumber
\end{align}
which together with (\ref{42}) gives us:
\begin{align}
\mathbb{P}\Big( \liminf_{t\rightarrow \infty}  \mathfrak{X}\big((t+1)Z_{H_0}\big) =\infty\Big)=1.
\end{align}
We can further conclude that:
\begin{align}
\mathbb{P}\Big( \liminf_{k\rightarrow \infty}  \mathfrak{X}\big(k\big) =\infty\Big)=1
\end{align}
since $ \mathbb{P}\big(\mathfrak{X}\big(k\big) \geq |2\alpha-1|^{Z_{H_0}}\mathfrak{X}\big(\lceil \frac{k}{Z_{H_0}} \big \rceil Z_{H_0} \big)\big)=1$ in view of (\ref{30}).

Therefore, for any integer  $H_0\geq1$, we have proved that belief divergence is achieved for all initial condition satisfying $\mathfrak{X}(0)>0$ if $\beta>\beta^\diamondsuit (H_0)$. Define
$$
\beta^\sharp:=\inf_{H_0\geq 1}\beta^\diamondsuit (H_0).
$$
With this choice of $\beta^\sharp$, the desired conclusion holds. \hfill$\square$

\subsection*{F. Proof of Proposition \ref{quasicub1}}
Note that there exist  $\{i_s,j_s\}\in \mathsf{G}_{\rm pst}$, $s=1,2,\dots, T$ with  $T\geq1$  such that
\begin{align}\label{19}
\mathrm{W}^+_{i_Tj_T}\cdots \mathrm{W}^+_{i_1j_1}=U
\end{align} if and only if for any $y(0)=y^0=(y_1^0 \dots y_n^0)'$, the dynamical system
\begin{align}\label{100}
y(k)=  \mathrm{W}^+_{i_kj_k} y(k-1), \ k=1,\dots,T
\end{align}
drives $y(k)=(y_1(k),\dots, y_n(k))'$ to $y(T)={\rm ave}(y(0)) \mathbf{1}$ where ${\rm ave}({y(0)})={\sum_{i=1}^n y_i^0}/{n}$. Thus we may study the matrix equality (\ref{19}) through individual node dynamics, which we leverage in the proof.

The claim follows from an induction argument. Assume that the desired  sequence of node pairs with length $\mathrm{T}_k=k2^{k-1}$ exists for $m=k$. Assume that $\mathsf{G}_{\rm pst}$ has a subgraph isomorphic to an $m+1$ dimensional hypercube.  Without loss of generality we assume $\mathsf{V}$ has been rewritten as $\{0,1\}^{k+1}$ following the definition of hypercube.

Now define $$
  \mathsf{V}^\dag_{0}:=\{i_1\times\dots\times i_{k+1} \in \mathsf{V}: i_{k+1}=0 \};\ \   \mathsf{V}^\dag_{1}:=\{i_1\times\dots\times i_{k+1} \in \mathsf{V}: i_{k+1}=1 \}.
  $$
It is easy to see that each of the subgraphs  $\mathsf{G}_{\mathsf{V}^\dag_{0}}$ and  $\mathsf{G}_{\mathsf{V}^\dag_{1}}$ contains a positive subgraph isomorphic with   an $m$-dimensional hypercube. Therefore, for any initial value of $y(0)$, the nodes in each set $\mathsf{G}_{\mathsf{V}^\dag_{s}},s=0,1$ can reach the same value, say $C^0(y(0))$ and $C^1(y(0))$, respectively. Then we select the following $2^k$ edges  for updates from $\mathsf{G}$:
 $$
 \{i_1\times\dots\times i_{k}\times 0, i_1\times\dots\times i_{k}\times 1\}: i_s\in\{0,1\}, s=1,\dots,k.
 $$
After these updates, all nodes reach the same value $(C^0(y(0))+C^1(y(0)))/2$ which has to be ${\rm ave}({y(0)})$ since the sum of the node beliefs is constant during this process. Thus, the desired  sequence of node pairs exists also for $m=k+1$, with a length
$$
\mathrm{T}_{k+1}=2\mathrm{T}_k +2^k=2k2^{k-1}+2^k=(k+1)2^{k}.
$$
This proves the desired conclusion.

\subsection*{G. Proof of Proposition \ref{quasicub2}}
The requirement of $\alpha=1/2$ is obvious since otherwise $\mathrm{W}^+_{ij}$ is nonsingular for all $\{i,j\}\in \mathsf{E}_{\rm pst}$, while ${\rm rank} U=1$. The necessity of $m=2^k$ for some $k\geq 0$ was proved in \cite{shi12} through an elementary number theory argument by constructing a particular initial value for which finite-time convergence can never be possible by pairwise averaging.

It remains to show that  $\mathsf{G}_{\rm pst}$ has a perfect matching.  Now suppose Eq. (\ref{19}) holds.
Without loss of generality we assume that Eq. (\ref{19}) is {\it irreducible}
in the sense that the equality  will no longer hold if any (one or more) matrices are removed from that sequence. The idea of the proof  is to analyze the dynamical system (\ref{100}) backwards from the final step. In this way we will recover a perfect matching from
$\big\{\{i_1,j_1\}, \dots, \{i_{T},j_{T}\}\big\}$. We divide the remaining of the proof into three steps.

\noindent {Step 1.} We first establish some property associated with $\{i_{T},j_{T}\}$. After the last step in (\ref{100}), two nodes $i_T$ and $j_T$ reach the same value, ${\rm ave}(y^0)$, along with all the other nodes. We can consequently  write
$$
y_{i_T} (T-1)={\rm ave}(y^0)+ h_T(y^0),\ \ y_{j_T} (T-1)={\rm ave}(y^0)- h_T(y^0),
$$
where $h_T(\cdot)$ is a real-valued function marking the error between $y_{i_T} (T-1)$, $y_{j_T} (T-1)$ and the true average ${\rm ave}(y^0)$.

Indeed, the set $\{y^0:h_T(y^0)=0\}$ is  explicitly given by
$$
\Big\{ y^0:(0\dots \underbrace{1}_{\mbox{$i_T$'th} } \dots \underbrace{-1}_{\mbox{$j_T$'th}} \dots 0) \mathrm{W}^+_{i_{T-1}j_{T-1}}\cdots \mathrm{W}^+_{i_1j_1} y^0=0\Big\},
$$
which is a linear subspace with dimension $n-1$ (recall that the equation $\mathrm{W}^+_{i_{T}j_{T}}\dots \mathrm{W}^+_{i_1j_1}=U$ is irreducible). Thus there must be $h_T(y^0)\neq 0$ for some initial value $y^0$.

\noindent {Step 2.} If there are only two nodes in the network, we are done. Otherwise $\{i_{T-1},j_{T-1}\}\neq\{i_T,j_T\}$. We make the following claim.

{\it Claim.} $i_{T-1},j_{T-1}\notin\{i_T,j_T\}$.

Suppose without loss of generality that $i_{T-1}=i_T$. Then
$$
y_{j_{T-1}}(T)=y_{j_{T-1}}(T-1)=y_{i_{T-1}}(T-1)=y_{i_{T}}(T-1)={\rm ave}(y^0)+ h_T(y^0).
$$
While on the other hand $y_{j_{T-1}}(T)={\rm ave}(y^0)$ for all $y^0$. The claim holds observing that as we just established, $\{h_T(y^0)\neq 0\}$ is a nonempty set.

We then write:
$$
y_{i_{T-1}} (T-2)={\rm ave}(y^0)+ h_{T-1}(y^0),\ \ y_{j_{T-1}} (T-2)={\rm ave}(y^0)- h_{T-1}(y^0)
$$
where $h_{T-1}(\cdot)$ is again a real-valued function and $h_{T-1}(y^0)\neq  0$ for some initial value $y^0$ (applying the same argument as for $h_{T}(y^0)\neq  0$). Note that
$$
\big \{y^0:h_{T}(y^0)\neq 0 \big \}\cap \big \{y^0:h_{T-1}(y^0)\neq 0 \big \}\ =\ \Big( \big \{y^0:h_{T}(y^0)= 0 \big \}\cup \big \{y^0:h_{T-1}(y^0)= 0 \big \}\Big)^c
$$
is  nonempty because  it is the complement of the union  of two  linear subspaces of dimension $n-1$ in $\mathds{R}^n$.

\noindent {Step 3.} Again, if there are only four nodes in the network, we are done. Otherwise, we can define:
\begin{align}
T_\star:= \max \Big\{\tau:  \{i_{\tau},j_\tau\}\nsubseteq\{i_{T-1},j_{T-1},i_T,j_T\} \Big\}
\end{align}
We emphasize  that $T_\star$ must exist since Eq. (\ref{19}) holds. As before, we have  $i_{T_\star},j_{T_\star}\notin\{i_{T-1},j_{T-1},i_T,j_T\}$ and $h_{T_\star}(y^0)$ can be found with $\{h_{T_\star}(y^0)=0\}$ being another $(n-1)$-dimensional subspace such that
$$
y_{i_{T_\star}} (T_\star-1)={\rm ave}(y^0)+ h_{T_\star}(y^0),\ \ y_{j_{T_\star}} (T_\star-1)={\rm ave}(y^0)- h_{T_\star}(y^0).
$$

We thus conclude that this argument can be proceeded recursively until we have found  a perfect matching  of  $\mathsf{G}_{\rm pst}$ in $\big\{\{i_1,j_1\}, \dots, \{i_{T},j_{T}\}\big\}$. We have now completed the proof.

\subsection*{H. Proof of Theorem \ref{strongbalance}}

We first state and prove intermediate lemmas that will be useful for the proofs of Theorems \ref{strongbalance}, \ref{thmweakbalance}, and \ref{thmergodic}.

\begin{lemma}\label{lemleader}
Assume that $\alpha\in (0,1)$. Let $i_1 \dots i_k$ be a path in the positive graph, i.e., $\{i_s,i_{s+1}\} \in \mathsf{G}_{\rm pst},s=1,\dots,k-1$.  Take a node $i_\ast \in \{i_1,\dots, i_k\}$. Then for any $\varepsilon>0$, there always exists an integer $Z_\star(\varepsilon)\geq 1$, such that we can select a sequence of node pairs  from $\{i_s,i_{s+1}\},s=1,\dots,k-1$ under  asymmetric updates  which guarantees
$$
J_{i_\ast i_s}(Z_\star) \leq 2A  \varepsilon,\ s\in  \{1,\dots, k\}
$$
  for all initial condition $x_{i_s}(0),s=1,\dots,k$.
\end{lemma}
{\it Proof.} The proof is easy and an appropriate sequence of node pairs can be built just observing that $J_{i_\ast i_s}\leq 2A$ for all $s\in  \{1,\dots, k\}$. \hfill$\square$

\begin{figure}[t]
\begin{center}
\includegraphics[height=2.4in]{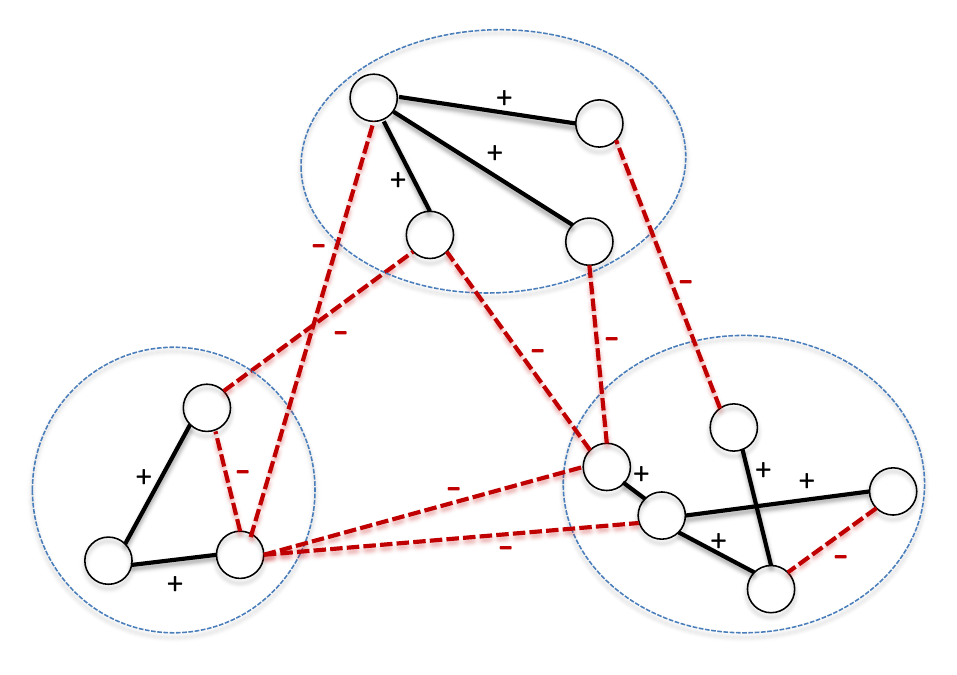}
\caption{There is a unique partition of $\mathsf{G}$ into subgraphs following the connected components of $\mathsf {G}_{\rm pst}$. Viewing each subgraph as a super node, the graph  is connected, and has only negative edges.} \label{partition}
\end{center}
\end{figure}

\begin{lemma}\label{lemsep} Fix $\alpha\in (0, 1) $ with $\alpha\neq 1/2$. Under belief dynamics (\ref{121}), there exist an integer $Z_0\geq 1$ and a constant $\vartheta_0 >0$ such that
\begin{align}\label{72}
\mathbb{P} \big( \exists \{i_\ast,j_\ast\} \in \mathsf{G}_{\rm neg}\ \mbox{s.t.}\  {J}_{i_\ast j_\ast}(Z_0)\geq \frac{1}{2n} \mathfrak{X}(0)\big) \geq \vartheta_0.
\end{align}
\end{lemma}
{\it Proof.}  We can always uniquely divide $\mathsf{V}$ into $m_0\geq 1$ mutually disjoint sets ${V}_1,\dots,{V}_{m_0}$ such that $\mathsf {G}_{\rm pst} ({{V}_k}), k=1,\dots, m_0$ are connected graphs, where $\mathsf {G}_{\rm pst}( {{V}_k})$ is the induced graph of $\mathsf {G}_{\rm pst}$ by node set ${V}_k$. The idea is to treat each $\mathsf {G}_{\rm pst}({{V}_k})$ as a {\it super node} (an illustration of this partition is shown in Figure \ref{partition}). Since  $\mathsf {G}$ is connected and $\mathsf {G}_{\rm neg}$ is nonempty, these super nodes form a connected graph whose edges are negative.

One can readily show that there exist two distinct  nodes $\eta_1,\eta_2 \in\mathsf{V}$ with $\eta_i\in {V}_{\nu_i},i=1,2$  (${V}_{\nu_1}$ and ${V}_{\nu_2}$ can be the same, of course) such that there is at least one {\it negative} edge between ${V}_{\nu_1}$ and ${V}_{\nu_2}$ and such that:
\begin{align}\label{70}
J_{\eta_1 \eta_2}(0)\geq \frac{1}{m_0}\mathfrak{X}(0).
 \end{align}
Now select $\upsilon_1 \in {V}_{\nu_1}$ and $\upsilon_2 \in {V}_{\nu_2}$ such that $\{\upsilon_1,\upsilon_2\}\in \mathsf{E}_{\rm neg}$. In view of Lemma \ref{lemleader} and observing that  asymmetric updates happen with a strictly positive probability, we can always find $\vartheta_0>0$ and $Z_0\geq 1$ (both functions of $(\alpha,n,a,b,c)$) such that:
\begin{align}\label{71}
 \mathbb{P} \Big( x_{\nu_i}(Z_0)=x_{\nu_i}(0),\ {J}_{\upsilon_i \nu_i}(Z_0)\leq \frac{1}{4n} \mathfrak{X}(0),\ i=1,2 \Big) \geq \vartheta_0,
\end{align}
(because $\mathsf {G}_{\rm pst}({{V}_{\nu_i}}),i=1,2$ are connected graphs). (\ref{72}) follows from (\ref{70}) and (\ref{71}) since $m_0\leq n$. \hfill$\square$

\begin{lemma}\label{lembounddivergence}
Fix $\alpha\in (0, 1) $ with $\alpha\neq 1/2$. Under belief dynamics (\ref{121}), there exists $\beta^\diamond(\alpha) >0$ such that  $\mathbb{P}(\limsup_{k\rightarrow \infty} \mathfrak{X}(k)=2A)=1$ for almost all initial beliefs  if  $ \beta>\beta^\diamond$.
\end{lemma}
{\it Proof.} In view of Lemma \ref{lemsep}, we have:
\begin{align}\label{73}
 \mathbb{P} \Big(  \mathfrak{X}(Z_0+t) \geq  \min \big\{\frac{(\beta+1)^t}{2n}\mathfrak{X}(0), 2A \big\} \Big)\geq \big(\frac{c p_\ast}{n}\big)^t \vartheta_0,\ \ t=0,1,\dots.
\end{align}
We can conclude that:
\begin{align}\label{74}
\mathbb{P}(\limsup_{k\rightarrow \infty} \mathfrak{X}(k)=2A)+\mathbb{P}(\limsup_{k\rightarrow \infty} \mathfrak{X}(k)=0)=1
\end{align}
as long as $\beta>0$ using the same argument as that used in the proof of statement (i) in Lemma \ref{thmliveordie}.

With (\ref{73}), we have:
\begin{align}
\mathbb{P} \big(  \mathfrak{X}(Z_0+1) \geq  \frac{\beta+1}{2n}\mathfrak{X}(0)  \big)\geq \frac{c p_\ast}{n} \vartheta_0
\end{align}
conditioned on  $\mathfrak{X}(0) \leq {4An}/{(1+\beta)}$. Moreover, (\ref{30}) still holds for belief dynamics (\ref{121}). Therefore, we can invoke exactly the same argument
as that used in the proof of Lemma \ref{lemdivergence} to conclude that there exists $\beta^\diamond(\alpha) >0$ such that
\begin{align}\label{75}
\mathbb{P}\big(\limsup_{k\rightarrow \infty} \mathfrak{X}(k)\geq {4An}/(1+\beta)\big )=1
\end{align}
for all $\beta>\beta^\diamond(\alpha)$. Combining (\ref{74}) and (\ref{75}), we get the desired result. \hfill$\square$

\begin{lemma}\label{lemsepevent} Assume that the graph is strongly balanced under partition $\mathsf{V}=\mathsf{V}_1\cup \mathsf{V}_2$, and that $\mathsf{G}_{\mathsf{V}_1}$ and $\mathsf{G}_{\mathsf{V}_2}$ are connected. Let $\alpha\in (0,1)\setminus\{ 1/2\}$. Fix the initial beliefs $x^0$. Then under belief dynamics (\ref{121}), there are two random variables, $B_1^\dag(x^0),B_2^\dag(x^0)$ both taking value in $\{-A,A\}$, such that
\begin{align}
\mathbb{P}\Big( \lim_{k\rightarrow \infty} x_i(k)=B_1^\dag,\ i\in \mathsf{V}_1; \lim_{k\rightarrow \infty} x_i(k)=B_2^\dag,\ i\in \mathsf{V}_2\Big| \mathcal{E}_{\rm sep}(\epsilon)\Big)=1\
\end{align}
for all $\epsilon>0$,
where by definition, $\mathcal{E}_{\rm sep}(\epsilon)$ is the $\epsilon$-separation event:
$$
\mathcal{E}_{\rm sep}(\epsilon):= \Big\{ \limsup_{k\rightarrow \infty} \max_{i\in \mathsf{V}_1,j\in \mathsf{V}_2}\big|x_{i}(k)-{x}_{j}(k)\big| \geq  \epsilon \Big\}.
$$
\end{lemma}
{\it Proof.} Suppose $x_{i_1}(0)-x_{i_2}(0) \geq \epsilon >0$ for $i_1\in\mathsf{V}_1$ and $i_2\in\mathsf{V}_2$. By assumption, $\mathsf{G}_{\mathsf{V}_1}$ and $\mathsf{G}_{\mathsf{V}_2}$ are connected. Thus, from Lemma \ref{lemleader}, there exist an integer  $Z_1\geq 1$ and a constant $\bar{p}$ (both depending on $\epsilon,n,\alpha,a,b$) such that
\begin{align}\label{76}
\min_{i\in \mathsf{V}_1 }x_{i}(Z_1) -\max_{i\in \mathsf{V}_2 }x_{i}(Z_1) \geq \frac{\epsilon}{2}
\end{align}
happens with probability at least $\bar{p}$. Intuitively Eq. (\ref{76}) characterizes the event where the beliefs in the two sets $\mathsf{V}_1$ and $\mathsf{V}_2$ are {\it completely separated}. Since all edges between the two sets are negative, conditioned on event (\ref{76}), it is then straightforward to see that almost surely we have $\lim_{k\rightarrow \infty} x_i(k) =A, i\in \mathsf{V}_1$ and $\lim_{k\rightarrow \infty} x_i(k) =-A, i\in \mathsf{V}_2$.

Given $\mathcal{E}_{\rm sep} (\epsilon)$, $\big \{\exists i_1\in\mathsf{V}_1, i_2\in\mathsf{V}_2\ \mbox{s.t.}\ x_{i_1}(k)-x_{i_2}(k) \geq \epsilon\ \mbox{for infinitely many $k$}\big\}$ is an almost sure event. Based on our previous discussion and by a simple stopping time argument, the Borel-Cantelli Lemma implies that the complete separation event happens almost surely given $\mathcal{E}_{\rm sep} (\epsilon)$. This completes the proof. \hfill$\square$

\begin{lemma}\label{lema/2}
Assume that the graph is strongly balanced under partition $\mathsf{V}=\mathsf{V}_1\cup \mathsf{V}_2$, and that $\mathsf{G}_{\mathsf{V}_1}$ and $\mathsf{G}_{\mathsf{V}_2}$ are connected. Suppose  $\alpha\in (0,1)\setminus\{ 1/2\}$. Then under dynamics (\ref{121}), there exists $\beta$ sufficiently large such that $\mathbb{P}\big (\mathcal{E}_{\rm sep}({A}/{2})\big)=1$ for almost all initial beliefs.
\end{lemma}
{\it Proof.} Let us first focus on a fixed time instant $k$.  Suppose $x_i(k)-x_j(k) \geq A$ for some $i,j\in \mathsf{V}$. If $i$ and $j$ belong to different sets $\mathsf{V}_1$ and  $\mathsf{V}_2$, we already have $\max_{i\in \mathsf{V}_1,j\in \mathsf{V}_2}\big|x_{i}(k)-{x}_{j}(k) \big| \geq A$. Otherwise, say $i,j\in \mathsf{V}_1$. There must be another node $l\in \mathsf{V}_2$. We have
 $\max_{i\in \mathsf{V}_1,j\in \mathsf{V}_2}\big|x_{i}(k)-{x}_{j}(k) \big| \geq A/2$ since either $|x_i(k)-x_l(k)| \geq A/2$ or $|x_j(k)-x_l(k)| \geq A/2$ must hold.  Therefore, we conclude that
 \begin{align}\label{136}
 \mathfrak{X}(k)\geq A\  \Longrightarrow \ \max_{i\in \mathsf{V}_1,j\in \mathsf{V}_2} \big|x_{i}(k)-{x}_{j}(k) \big| \geq A/2.
 \end{align}

Then the  desired conclusion  follows directly from Lemma \ref{lembounddivergence}. \hfill$\square$

Theorem \ref{strongbalance} is a direct consequence of Lemmas \ref{lemsepevent} and \ref{lema/2}.

\subsection*{I. \ Proof of Theorem \ref{thmweakbalance}}

The proof is similar to that of Theorem \ref{strongbalance}. We just provide the main arguments.

First by Lemma \ref{lembounddivergence} we have $\mathbb{P}(\limsup_{k\rightarrow \infty} \mathfrak{X}(k)=2A)=1$ for almost all initial values with sufficiently large $\beta$. Then as for (\ref{136}), we have
 \begin{align}
 \mathfrak{X}(k)\geq A\  \Longrightarrow \ \max_{i\in \mathsf{V}_s,j\in \mathsf{V}_t, s\neq t\in\{1,\dots,m\}}\big|x_{i}(k)-{x}_{j}(k)\big|\geq  \frac{A}{m},
 \end{align}
 where $m\geq 2$ comes from the definition of weak balance.  Therefore, introducing
$$
\mathcal{E}_{\rm sep}^\ast(\epsilon):= \Big\{ \limsup_{k\rightarrow \infty} \max_{i\in \mathsf{V}_s,j\in \mathsf{V}_t,s\neq t\in\{1,\dots,m\}}\big|x_{i}(k)-{x}_{j}(k)\big|\geq \epsilon \Big\},
$$
we can show that $\mathbb{P}\big (\mathcal{E}_{\rm sep}^\ast({A}/{m})\big)=1$ for almost all initial beliefs, for sufficiently large $\beta$.

Next, suppose there exist a constant $\eta>0$ and two node sets $\mathsf{V}_{i_1}$ and $\mathsf{V}_{i_2}$ with $i_1,i_2\in\{1,\dots,m\}$ such that the complete separation event
\begin{align}\label{137}
\min_{i\in \mathsf{V}_{i_1}}x_{i}(k) -\max_{i\in \mathsf{V}_{i_2} }x_{i}(k) \geq \eta
\end{align}
happens. Recall that the underlying graph is complete. Then if $(\beta+1)\eta \geq 2A$, we can always select  $Z_\ast:=|\mathsf{V}_{i_1}|+|\mathsf{V}_{i_2}|$ negative edges between nodes in the sets $\mathsf{V}_{i_1}$ and $\mathsf{V}_{i_2}$, so that after the corresponding updates:
\begin{align}\label{80}
x_i(k+Z_\ast)=A, i\in\mathsf{V}_{i_1},\ \  x_i(k+Z_\ast)=-A, i\in\mathsf{V}_{i_2}.
\end{align}
One can easily see that we can continue to build the (finite) sequence of edges for updates, such that nodes in $\mathsf{V}_{k}$ will hold the same belief in $\{-A,A\}$, for all $k=1,\dots,m$. After this sequence of updates, the beliefs held at the various nodes remain unchanged (two nodes with the same belief cannot influence each other, even in presence of a negative link; and two nodes with different beliefs are necessarily enemies). To summarize, conditioned on the complete separation event (\ref{137}), we can select a sequence of node pairs under which belief clustering is reached, and this clustering state is an {\it absorbing} state.

Finally, the Borel-Cantelli Lemma and  $\mathbb{P}\big (\mathcal{E}_{\rm sep}^\ast({A}/{m})\big)=1$ guarantee that almost surely the complete separation event (\ref{137}) happens an infinite number of times if $\eta={A}/{2m}$ in view of Lemma \ref{lemleader}. The end of the proof is then done as in that of Theorem \ref{strongbalance}.

\subsection*{J. \ Proof of Theorem \ref{thmergodic}}

Again the result is obtained by combining Lemmas \ref{lemleader} and \ref{lembounddivergence} with Borel-Cantelli lemma.

\vspace{10mm}

\section*{Acknowledgement}
This work has been supported in part
by the Knut and Alice Wallenberg Foundation, the Swedish Research
Council, and KTH SRA TNG. The authors gratefully thank Dr. Shuangshuang Fu for her generous help in preparing the numerical examples.

\medskip

\medskip

\noindent {\sc  \large Guodong Shi} \\
\noindent {\small Research School of Engineering,  College of Engineering and Computer Science,  \\
The Australian National University,  Canberra ACT 0200, Australia}\\  {\small Email: } {\tt\small guodong.shi@anu.edu.au}

\medskip

 \medskip

\medskip

\noindent {\sc \large  Alexandre Proutiere, Mikael Johansson, and Karl H. Johansson} \\
\noindent {\small  ACCESS Linnaeus Centre,
   School of Electrical Engineering,
KTH Royal Institute of Technology,
\\ Stockholm 100 44, Sweden }\\
       {\small Email: 』 {\tt\small alepro@kth.se, mikaelj@kth.se, kallej@kth.se}

\medskip

\medskip

\noindent {\sc  \large John S. Baras} \\
\noindent {\small Institute for Systems Research, University of Maryland,\\
College Park, MD 20742, USA}\\
{\small Email: } {\tt\small baras@umd.edu}

\end{document}